\documentclass{emulateapj}
\usepackage{epsfig,natbib}
\usepackage{graphicx}
\usepackage{enumerate}
\usepackage{amsmath}
\usepackage{multirow}
\usepackage{bigdelim}
\usepackage{longtable}
\usepackage{threeparttable}
\usepackage{subfigure}
\usepackage{textcomp}
\usepackage{gensymb}
\usepackage{color}
\usepackage{morefloats}
\usepackage{xspace}
\usepackage[T1]{fontenc}
\usepackage{placeins}

\newcommand{\thisstudy}{This study}

\newcommand{\baistarteff}{\ensuremath{5887 \pm 46\xspace}}
\newcommand{\baistarlogg}{\ensuremath{4.03 \pm 0.05\xspace}}
\newcommand{\baistarfe}{\ensuremath{-0.047 \pm 0.02\xspace}}
\newcommand{\baistarvsini}{\ensuremath{3.7 \pm 2.0\xspace}}
\newcommand{\baistarmass}{\ensuremath{1.11 \pm 0.04\xspace}}
\newcommand{\baistarrad}{\ensuremath{1.67 \pm 0.12\xspace}}
\newcommand{\baistarrphk}{\ensuremath{-5.27}\xspace}
\newcommand{\baistarshk}{\ensuremath{0.128}\xspace}
\newcommand{\baistarvmag}{\ensuremath{11.710 \pm 0.186}\xspace}

\newcommand{\hbcstarteff}{\ensuremath{5496 \pm 46\xspace}}
\newcommand{\hbcstarlogg}{\ensuremath{4.42 \pm 0.05\xspace}}
\newcommand{\hbcstarfe}{\ensuremath{0.06 \pm 0.03\xspace}}
\newcommand{\hbcstarvsini}{\ensuremath{< 2.0}\xspace}
\newcommand{\hbcstarmass}{\ensuremath{0.92 \pm 0.03\xspace}}
\newcommand{\hbcstarrad}{\ensuremath{0.95 \pm 0.05\xspace}}
\newcommand{\hbcstarrphk}{\ensuremath{-5.15}\xspace}
\newcommand{\hbcstarshk}{\ensuremath{0.140}\xspace}
\newcommand{\hbcstarvmag}{\ensuremath{12.102 \pm 0.212}\xspace}

\newcommand{\baibtt}{\ensuremath{2455817.0092 \pm 0.0051\xspace}}
\newcommand{\baibper}{\ensuremath{5.06963 \pm 0.00081\xspace}}
\newcommand{\baibincdeg}{\ensuremath{86.6^{+2.4}_{-4.4}}\xspace}
\newcommand{\baibrprs}{\ensuremath{0.01353^{+0.00174}_{-0.00080}}\xspace}

\newcommand{\baibrsa}{\ensuremath{0.127^{+0.048}_{-0.013}}\xspace}
\newcommand{\baibuzero}{\ensuremath{0.52 \pm 0.01\xspace}}
\newcommand{\baibuone}{\ensuremath{0.19 \pm 0.01\xspace}}
\newcommand{\baibb}{\ensuremath{0.47 \pm 0.31\xspace}}
\newcommand{\baibtdur}{\ensuremath{4.71^{+0.45}_{-0.26}}\xspace}
\newcommand{\baibrp}{\ensuremath{2.49^{+0.34}_{-0.24}}\xspace}
\newcommand{\baibrhostar}{\ensuremath{0.36^{+0.14}_{-0.22}}\xspace}
\newcommand{\baiba}{\ensuremath{0.05983 \pm 0.00072\xspace}}
\newcommand{\baibfinc}{\ensuremath{840 \pm 125\xspace}}
\newcommand{\baibteq}{\ensuremath{1372 \pm 51\xspace}}
\newcommand{\baibjitj}{\ensuremath{5.0 \pm 0.8\xspace}}
\newcommand{\baibgammaj}{\ensuremath{-2.5 \pm 1.0\xspace}}
\newcommand{\baibk}{\ensuremath{7.4 \pm 1.2\xspace}}
\newcommand{\baibmp}{\ensuremath{21.3 \pm 3.6\xspace}}
\newcommand{\baibrhop}{\ensuremath{7.8 \pm 2.7\xspace}}

\newcommand{\hbcbtt}{\ensuremath{2456226.4368 \pm 0.0016\xspace}}
\newcommand{\hbcbper}{\ensuremath{0.571336 \pm 0.000020\xspace}}
\newcommand{\hbcbincdeg}{\ensuremath{80.2^{+7.0}_{-12.7}}\xspace}
\newcommand{\hbcbrprs}{\ensuremath{0.01745^{+0.00187}_{-0.00079}}\xspace}

\newcommand{\hbcbrsa}{\ensuremath{0.366^{+0.121}_{-0.036}}\xspace}
\newcommand{\hbcbuzero}{\ensuremath{0.459 \pm 0.001\xspace}}
\newcommand{\hbcbuone}{\ensuremath{0.225 \pm 0.001\xspace}}
\newcommand{\hbcbb}{\ensuremath{0.47 \pm 0.32\xspace}}
\newcommand{\hbcbtdur}{\ensuremath{1.79^{+0.56}_{-0.23}}\xspace}
\newcommand{\hbcbrp}{\ensuremath{1.82^{+0.20}_{-0.14}}\xspace}
\newcommand{\hbcbrhostar}{\ensuremath{1.18^{+0.43}_{-0.68}}\xspace}
\newcommand{\hbcba}{\ensuremath{0.01312 \pm 0.00014\xspace}}
\newcommand{\hbcbfinc}{\ensuremath{4293 \pm 483\xspace}}
\newcommand{\hbcbteq}{\ensuremath{2063 \pm 58\xspace}}
\newcommand{\hbcbgammaj}{\ensuremath{-2.2 \pm 1.0\xspace}}
\newcommand{\hbcbjitj}{\ensuremath{5.1 \pm 0.7\xspace}}
\newcommand{\hbcbk}{\ensuremath{7.2 \pm 1.3\xspace}}
\newcommand{\hbcbmp}{\ensuremath{9.0 \pm 1.6\xspace}}
\newcommand{\hbcbrhop}{\ensuremath{8.57^{+4.64}_{-2.80}}\xspace}

\newcommand{\hbcctt}{\ensuremath{2456238.7352 \pm 0.0042\xspace}}
\newcommand{\hbccper}{\ensuremath{13.3387 \pm 0.0018\xspace}}
\newcommand{\hbccincdeg}{\ensuremath{89.0^{+0.7}_{-1.4}}\xspace}
\newcommand{\hbccrprs}{\ensuremath{0.0265^{+0.0036}_{-0.0015}}\xspace}
\newcommand{\hbccrsa}{\ensuremath{0.0368^{+0.0159}_{-0.0041}}\xspace}
\newcommand{\hbccuzero}{\ensuremath{0.459 \pm 0.001\xspace}}
\newcommand{\hbccuone}{\ensuremath{0.225 \pm 0.001\xspace}}
\newcommand{\hbccb}{\ensuremath{0.47 \pm 0.32\xspace}}
\newcommand{\hbcctdur}{\ensuremath{3.50 \pm 0.21\xspace}}
\newcommand{\hbccrp}{\ensuremath{2.77^{+0.37}_{-0.23}}\xspace}
\newcommand{\hbccrhostar}{\ensuremath{2.13^{+0.92}_{-1.40}}\xspace}
\newcommand{\hbcca}{\ensuremath{0.1071 \pm 0.0015\xspace}}
\newcommand{\hbccfinc}{\ensuremath{64 \pm 7\xspace}}
\newcommand{\hbccteq}{\ensuremath{722 \pm 20\xspace}}
\newcommand{\hbccmp}{\ensuremath{5.7 \pm 6.1\xspace}}
\newcommand{\hbccrhop}{\ensuremath{1.3 \pm 1.6\xspace}}
\newcommand{\hbcck}{\ensuremath{1.6 \pm 1.7\xspace}}
\newcommand{\hbccthreesigk}{\ensuremath{6.7}\xspace}
\newcommand{\hbccthreesigmp}{\ensuremath{24.4}\xspace}

\makeatletter

\newcommand{\Rmnum}[1]{\expandafter\@slowromancap\romannumeral #1@}
\newcommand{\lsun}{\ensuremath{L_{\odot}}\xspace}
\newcommand{\msun}{\ensuremath{M_{\odot}}\xspace}
\newcommand{\rsun}{\ensuremath{R_{\odot}}\xspace}
\newcommand{\mearth}{\ensuremath{M_{\oplus}}\xspace}
\newcommand{\rearth}{\ensuremath{R_{\oplus}}\xspace}
\newcommand{\kms}{km~s$^{-1}$\xspace}

\newcommand{\mytilde}{\raise.17ex\hbox{$\scriptstyle\mathtt{\sim}$}}

\newcommand{\shk}{$S_\mathrm{HK}$\xspace}

\newcommand{\lrphk}{$\log R^\prime_\mathrm{HK}$\xspace}

\newcommand{\ms}{m\,s$^{-1}$\xspace}

\newcommand{\Mstar}{\ensuremath{M_{\star}}\xspace}
\newcommand{\Rstar}{\ensuremath{R_{\star}}\xspace} 
\newcommand{\Lstar}{\ensuremath{L_{\star}}\xspace} 
\newcommand{\fe}{$[$Fe/H$]$\xspace}
\newcommand{\Teff}{$T_{\mathrm{eff}}$\xspace}  
\newcommand{\logg}{\ensuremath{\log g}\xspace} 
\newcommand{\vsini}{\ensuremath{v \sin i}\xspace} 
\newcommand{\rhostar}{\ensuremath{\rho_{\star, \mathrm{circ}}}\xspace} 
\newcommand{\gcc}{g\,cm$^{-3}$\xspace} 
\newcommand{\dvdt}{\ensuremath{dv/dt}\xspace} 

\newcommand{\Mp}{\ensuremath{M_p}\xspace} 
\newcommand{\Rp}{\ensuremath{R_p}\xspace}

\newcommand{\rhop}{\ensuremath{\rho_p}\xspace}
 
\newcommand{\Kepler}{\textit{Kepler}\xspace} 
\newcommand{\ktwo}{\textit{K2}\xspace}

\newcommand{\sinc}{$S_{\mathrm{inc}}$\xspace}
\newcommand{\searth}{$S_{\oplus}$\xspace}

\newcommand{\rprs}{\ensuremath{R_p/R_\star}\xspace}
\newcommand{\rsa}{\ensuremath{R_\star/a}\xspace}
\newcommand{\ars}{\ensuremath{a/R_\star}\xspace}
\newcommand{\tdur}{\ensuremath{T_\mathrm{14}}\xspace}
\newcommand{\uzero}{\ensuremath{u_0}\xspace}
\newcommand{\uone}{\ensuremath{u_1}\xspace}

\newcommand{\bai}{EPIC\,206153219\xspace}
\newcommand{\ktwobai}{K2-66\xspace}
\newcommand{\hbc}{EPIC\,220674823\xspace}
\newcommand{\ktwohbc}{K2-106\xspace}
\newcommand{\menv}{$M_\mathrm{env}$\xspace}
\newcommand{\mcore}{$M_\mathrm{core}$\xspace}
\newcommand{\mrock}{$M_\mathrm{rock}$\xspace}
\newcommand{\dkp}{\ensuremath{\Delta{K_p}}\xspace} 

\newcommand{\howard}{H13\xspace} 
\newcommand{\pepe}{P13\xspace} 
\newcommand{\ojeda}{S13\xspace} 
\newcommand{\batalha}{B11\xspace}
\newcommand{\esteves}{E15\xspace} 
\newcommand{\valenti}{V05\xspace} 
\newcommand{\vonbraun}{V11\xspace}

\newcommand{\demory}{D16\xspace} 
\newcommand{\becker}{B15\xspace} 
 
\newcommand{\sinukoff}{S17\xspace} 
\newcommand{\haywood}{H14\xspace}
\newcommand{\bruntt}{B10\xspace}
\newcommand{\leger}{L09\xspace}

\makeatother

\shortauthors{Sinukoff}
\shorttitle{K2-66 and K2-106}

\begin{document}

\pagenumbering{arabic}

\title{\ktwobai{b} and \ktwohbc{b}:  Two extremely hot sub-Neptune-size planets with high densities}

\author{
Evan Sinukoff\altaffilmark{1,2,16}
Andrew W.\ Howard\altaffilmark{2}, 
Erik A.\ Petigura\altaffilmark{3,17}, 
Benjamin J.\ Fulton\altaffilmark{1,2,18}, 
Ian J.\ M.\ Crossfield\altaffilmark{4,19}, 
Howard Isaacson\altaffilmark{5},
Erica Gonzales\altaffilmark{6},
Justin R.\ Crepp\altaffilmark{6},
John M.\ Brewer\altaffilmark{7},
Lea Hirsch\altaffilmark{5}, 
Lauren M.\ Weiss\altaffilmark{8}, 
David R.\ Ciardi\altaffilmark{9},
Joshua E.\ Schlieder\altaffilmark{10},
Bjoern Benneke\altaffilmark{3}, 
Jessie L. Christiansen\altaffilmark{9},
Courtney D.\ Dressing\altaffilmark{3,17},
Brad M.\ S.\ Hansen\altaffilmark{11}, 
Heather A.\ Knutson\altaffilmark{3},
Molly Kosiarek\altaffilmark{4},
John H.\ Livingston\altaffilmark{12},
Thomas P.\ Greene\altaffilmark{13}
Leslie A.\ Rogers\altaffilmark{14}
S\'ebastien L\'epine\altaffilmark{15}
}




\altaffiltext{1}{Institute for Astronomy, University of Hawai`i at M\={a}noa, Honolulu, HI 96822, USA} 
\altaffiltext{2}{Cahill Center for Astrophysics, California Institute of Technology, 1216 East California Boulevard, Pasadena, CA 91125, USA}
\altaffiltext{3}{Division of Geological and Planetary Sciences, California Institute of Technology, 1255 East California Blvd, Pasadena, CA 91125, USA}
\altaffiltext{4}{Department of Astronomy \& Astrophysics, University of California Santa Cruz, 1156 High St., Santa Cruz, CA, USA}
\altaffiltext{5}{Astronomy Department, University of California, Berkeley, CA, USA}
\altaffiltext{6}{Department of Physics, University of Notre Dame, 225 Nieuwland Science Hall, Notre Dame, IN, USA}
\altaffiltext{7}{Department of Astronomy, Yale University and 260 Whitney Avenue, New Haven, CT 06511, USA}
\altaffiltext{8}{Institut de Recherche sur les Exoplan\`etes, D\`epartement de Physique, Universit\`e de Montr\`eal, C.P.\ 6128, Succ.\ Centre-ville, Montr\'eal, QC H3C 3J7, Canada}
\altaffiltext{9}{IPAC-NExScI, Mail Code 100-22, Caltech, 1200 E. California Blvd., Pasadena, CA 91125, USA}
\altaffiltext{10}{Exoplanets and Stellar Astrophysics Laboratory, NASA Goddard Space Flight Center, Greenbelt, MD 20771, USA}
\altaffiltext{11}{Department of Physics \& Astronomy and Institute of Geophysics \& Planetary Physics, University of California Los Angeles, Los Angeles, CA 90095, USA}
\altaffiltext{12}{Department of Astronomy, The University of Tokyo, 7-3-1 Bunkyo-ku, Tokyo 113-0033, Japan}
\altaffiltext{13}{NASA Ames Research Center, Space Science and Astrobiology Division, M.S. 245-6, Moffett Field, CA 94035, USA}
\altaffiltext{14}{Department of Astronomy \& Astrophysics, University of Chicago, 5640 South Ellis Avenue, Chicago, IL 60637, USA}
\altaffiltext{15}{Department of Physics and Astronomy, Georgia State University, GA, USA}


\altaffiltext{16}{NSERC Postgraduate Research Fellow}
\altaffiltext{17}{Hubble Fellow}
\altaffiltext{18}{NSF Graduate Research Fellow}
\altaffiltext{19}{NASA Sagan Fellow}

\begin{abstract}
We report precise mass and density measurements of two extremely hot sub-Neptune-size planets from the \ktwo mission using radial velocities, \ktwo photometry, and adaptive optics imaging.  \ktwobai harbors a close-in sub-Neptune-sized (\baibrp\,\rearth) planet (\ktwobai{b}) with a mass of \baibmp\,\mearth. Because the star is evolving up the sub-giant branch, \ktwobai{b} receives a high level of irradiation, roughly twice the main sequence value. \ktwobai{b} may reside within the so-called ``photoevaporation desert'', a domain of planet size and incident flux that is almost completely devoid of planets. Its mass and radius imply that \ktwobai{b} has, at most, a meager envelope fraction (<  5\%) and perhaps no envelope at all, making it one of the largest planets without a significant envelope.  \ktwohbc hosts an ultra-short-period planet ($P$ = 13.7 hrs) that is one of the hottest sub-Neptune-size planets discovered to date.  Its radius (\hbcbrp\,\rearth) and mass (\hbcbmp\,\mearth) are consistent with a rocky composition, as are all other small ultra-short-period planets with well-measured masses. \ktwohbc also hosts a larger, longer-period planet (\Rp = \hbccrp\,\rearth, $P$ = 13.3 days) with a mass less than \hbccthreesigmp\,\mearth at 99.7\% confidence. \ktwobai{b} and \ktwohbc{b} probe planetary physics in extreme radiation environments. Their high densities reflect the challenge of retaining a substantial gas envelope in such extreme environments.
\end{abstract}


\section{Introduction}

Approximately one third of Sun-like stars host planets between the size of Earth and Neptune (``sub-Neptunes'') with orbital periods $P$~<~100 days \citep{Howard12, Fressin13, Petigura13b, Burke15}.   Most sub-Neptunes detected to date were discovered by the prime \Kepler mission (2009--2013).   While \Kepler provided a detailed measure of the distribution of planet radii, only a few tens of stars hosting sub-Neptunes were bright enough for secure mass-measurements by current generation precision radial velocity (RV) facilities \citep[e.g.][]{Marcy14a}. Many other planets have masses measured from transit timing variations \citep[TTVs,][]{Holman05, Agol05}, a technique that is limited to compact, multiplanet systems \citep[e.g.][]{Carter12, Hadden14}.  

Mass and radius measurements yield planet densities, which can be used to infer bulk compositions and probe planet formation histories.  From the dozens of sub-Neptunes with measured densities, bulk compositional trends have become apparent.  Most notably, the majority of planets smaller than $\approx$ 1.6\,\rearth have primarily rocky compositions, whereas most larger planets have lower densities, consistent with the presence of extended envelopes of H/He and other low-density volatiles \citep{Weiss14, Marcy14a, Rogers15, Dressing15a}.  

This overall trend in bulk compositions likely has a temperature dependence, which has yet to be fully explored.  The gaseous envelopes of planets at extreme temperatures are subjected to photoevaporation by the incident radiation from their host stars \citep[e.g.][]{Owen13, Lopez14}.  Probing planets at extreme temperatures is crucial to understand these sculpting effects and the formation histories of planets close to their host stars.  If these planets did form as mini-Neptunes and/or giant planets, studying the masses and compositions of their remnants provides insight into the nature of the cores of such planets, specifically the mechanisms that formed them, put them so close to their host stars, and removed their surrounding envelopes.  

Recent studies of planet occurrence as a function of radius and temperature have shed light on the formation and evolution of sub-Neptunes.  The prime \Kepler mission revealed that the occurrence of 2--4\,\rearth planets drops significantly at very short orbital periods \citep[$P$< 10 days,][]{Howard12, Fressin13}.  Moreover, from a study of Kepler planets and planet candidates, including 157 with astroseismically characterized host stars, \citet{Lundkvist16} reported a complete absence of planets with radii 2.2--3.8 \rearth and incident fluxes \sinc$>$ 650 \searth.  Evolutionary models have explained this gap as a ``photoevaporation desert'', because planets in this size and temperature regime have their envelopes stripped by photoevaporation \citep{Owen13, Lopez13}.   Alternatively, smaller planet cores might form too late and/or too close to the star to accrete much gas and grow in size \citep{Lee16}.   

Another rare sub-class of small planets are those with orbital periods $P$ $<$ 1 day, known as ``ultra-short-period'' planets (hereafter USPs).   They exist around $\sim$ 1\% of Sun-like stars \citep{Ojeda14}.  While it is unclear how USPs form and how they end up so close to the star, there are several observational clues:  Systems with USPs commonly host additional planets, which might have played a role in their formation and/or migration histories.  Moreover, \citet{Ojeda14} measured a sharp decrease in the occurrence of USPs larger than $\sim$ 1.4\,\rearth, and a complete lack of USPs $>$2.0\,\rearth.   \citet{Lopez16} showed that the observed dearth of USPs \Rp = 2--4\,\rearth suggests that they formed with water-poor H/He envelopes that were subsequently lost via photoevaporation.  

Bulk density measurements of these two rare types of sub-Neptunes can reveal whether they are bare cores, or contain a significant amount of volatiles.  Unfortunately, there have been few opportunities to study their compositions.  The few of them discovered in the prime \Kepler field orbit stars too faint for spectroscopic follow-up.  However, in 2014, NASA's \ktwo mission began a new chapter in the search for planets orbiting bright stars.  The \Kepler spacecraft has been collecting precise photometry of numerous fields along the ecliptic plane, each for nearly three continuous months \citep{Howell14}.  With 10,000--20,000 stars per campaign,  hundreds of transiting planet candidates have been discovered \citep{Vanderburg15, Pope16, Barros16, Adams16pig}, many of which have been statistically validated or confirmed as planets \citep{Sinukoff16, Crossfield16}.  This includes several USPs around bright stars amenable to Doppler spectroscopy, including WASP-47e \citep{Becker15, Dai15, Sinukoff17a} and HD 3167b \citep{Vanderburg16}.  \ktwo also provides an opportunity to probe the compositions of planets in and at the boundaries of the photoevaporation desert.

Here we report the first mass and density measurements of a planet in the photoevaporation desert as well as the mass and density of a USP planet in a multiplanet system.  \ktwobai (\bai) is a G1 subgiant star in \ktwo Campaign 3 (C3), which hosts a transiting sub-Neptune in the photoevaporation desert.   \ktwohbc (\hbc) is a G-star in \ktwo Campaign 8 (C8) with two transiting sub-Neptunes, including a USP sub-Neptune (\ktwohbc{b}).  We note that \ktwobai{b} was first reported as a planet candidate by \citet{Vanderburg15} and statistically validated by \citet{Crossfield16}.  Both \ktwohbc planets were first reported and statistically validated by \citet{Adams16} as part of the Short-Period Planets Group effort (SuPerPiG).  

In \S\ref{sec:obs} we describe the methods by which we generate stellar light curves from raw \ktwo photometry and summarize our adaptive optics imaging and Doppler observations.  \S\ref{sec:analysis} explains our analysis of the resulting light curves, AO images, and RV time-series to precisely characterize the host stars and determine planet masses and radii. In \S\ref{sec:discussion}, we present our results, discuss possible planet compositions, and place these planets in context with other sub-Neptunes.  Concluding statements are provided in \S\ref{sec:conclusion}.




\section{Observations}
\label{sec:obs}

\subsection{\ktwo Photometry}
NASA's \Kepler Telescope collected nearly continuous photometry of \ktwobai from 2014 November 15 -- 2015 January 23 UT (69 days) as part of \ktwo Campaign 3.  \ktwohbc was observed from 2016 January 04 -- 2016 March 23 UT (80 days) as part of \ktwo Campaign 8. We generated stellar light curves from the respective target pixel files using the same procedures detailed in \citet{Sinukoff16} and \citet{Crossfield16}.  The same Gaussian process was used to model and subtract the spacecraft motion from \ktwo pixel data.  We use the same K2-66 light curve presented in \citet{Crossfield16}, so we do not display it in this work. 

\subsection{Adaptive Optics Imaging}

We observed \ktwohbc on 2016 August 24 UT with the high-contrast adaptive optics (AO) system on the Keck-II telescope using the NIRC2 imaging instrument (PI: Keith Matthews). The images were obtained in the narrow camera mode using a 3-point dither pattern with nods of $2\arcsec$ in each cardinal direction to remove background light.  The K$_{s}$ filter was used for all observations.  Conditions were foggy and the star was at airmass 1.2 with seeing of $0.\arcsec8$ during the observations.  \citet{Crossfield16} presented NIRC2 adaptive optics imaging of \ktwobai obtained by our group, which we do not show here. The star was found to be single.  Moreover, \citet{Adams16} presented similar NIRC2 observations of \ktwohbc, finding no evidence of secondary sources.

\subsection{Radial Velocity Measurements}
RV measurements of \ktwobai and \ktwohbc were made using HIRES \citep{Vogt94} at the W.\ M.\ Keck Observatory.  We collected 38 RV measurements of \ktwobai from 2015 September 20 UT to 2017 January 07 UT and 35 RV measurements of \ktwohbc from 2016 August 12 UT to 2017 January 22 UT.  Observations and data reduction followed the usual methods of the California Planet Search \citep[CPS;][]{Howard10}.  An iodine cell was used for each observation as a wavelength calibrator and point spread function (PSF) reference.  The ``C2'' decker ($0\farcs87$ $\times$ 14\arcsec{} slit) provided spectral resolution $R$ $\approx$ 55,000 and allowed for the sky background to be measured and subtracted.  An exposure meter was used to automatically terminate exposures after reaching a target signal-to-noise ratio (SNR) per pixel at 550\,nm.  Most \ktwobai exposures were terminated at SNR $\approx$ 100 and typically lasted 20 min. \ktwohbc exposures proceeded until SNR $\approx$ 125 ($\sim$ 25 min).  For each star, a single iodine-free exposure was taken at roughly twice the SNR using the ``B3''  decker ($0\farcs57$ $\times$ 14\arcsec{} slit).  The standard CPS Doppler pipeline was used to measure RVs \citep{Marcy92, Valenti95, Butler96, Howard09}.   RV measurements are listed in Tables \ref{tb:RVdata_bai} and \ref{tb:RVdata_hbc} for \ktwobai and \ktwohbc, respectively.

\begin{deluxetable}{lrrr}[h!]
\tablecaption{\ktwobai Relative radial velocities, Keck-HIRES}
\tablehead{\colhead{BJD} & \colhead{RV} [\ms] & \colhead{Unc. [\ms]\tablenotemark{a}} & \colhead{\shk}\tablenotemark{b}}
\startdata
2457286.044784 & 6.58 & 4.19 & N/A \\
2457580.106140 & 7.14 & 2.24 & 0.127 \\
2457583.113840 & -13.31 & 2.18 & 0.127 \\
2457585.922824 & 3.35 & 2.10 & 0.128 \\
2457586.022505 & 2.63 & 2.17 & 0.128 \\
2457586.073226 & 4.15 & 2.20 & 0.127 \\
2457587.027388 & -8.39 & 2.06 & 0.129 \\
2457588.028820 & -2.36 & 2.07 & 0.128 \\
2457595.974851 & 9.66 & 2.63 & 0.116 \\
2457596.997324 & -3.35 & 4.25 & N/A \\
2457599.015841 & -7.50 & 2.21 & 0.125 \\
2457600.041053 & -0.06 & 1.99 & 0.128 \\
2457601.008159 & 5.21 & 2.29 & 0.126 \\
2457612.841886 & -8.28 & 2.64 & 0.128 \\
2457613.983431 & -16.35 & 2.69 & 0.131 \\
2457615.860156 & 6.51 & 2.99 & 0.133 \\
2457616.885444 & 13.79 & 2.84 & 0.130 \\
2457622.027461 & 14.11 & 3.09 & 0.126 \\
2457622.093780 & -1.18 & 3.37 & 0.125 \\
2457651.964266 & 2.48 & 2.83 & 0.135 \\
2457652.025942 & 9.88 & 2.80 & 0.128 \\
2457652.937923 & -9.59 & 2.88 & 0.133 \\
2457653.926554 & -7.16 & 2.76 & 0.136 \\
2457653.968022 & -7.98 & 2.67 & 0.135 \\
2457668.732792 & -0.47 & 2.71 & 0.118 \\
2457678.880082 & -1.04 & 3.05 & 0.125 \\
2457679.758736 & -0.64 & 2.65 & 0.130 \\
2457697.840632 & -2.61 & 2.70 & 0.124 \\
2457711.713727 & -3.43 & 2.79 & 0.124 \\
2457712.717828 & 0.82 & 2.66 & 0.127 \\
2457713.715934 & 5.43 & 2.68 & 0.129 \\
2457714.779542 & -8.83 & 3.03 & 0.128 \\
2457716.765754 & 2.09 & 2.95 & 0.125 \\
2457745.716553 & -15.72 & 2.80 & 0.127 \\
2457745.763482 & -24.08 & 5.15 & N/A \\
2457746.704085 & -1.96 & 2.73 & 0.128 \\
2457747.720099 & -6.56 & 2.59 & 0.127 \\
2457760.710967 & -6.37 & 3.04 & 0.124
\enddata
\tablenotetext{a}{Uncertainties estimated from the dispersion in the radial velocity measured from 718 chunks. These uncertainties do not include ``jitter'' which is incorporated as a free parameter during the RV modeling ($\sigma_{\rm jit}$, Table \ref{tb:params_bai}).} 
\tablenotetext{b}{For three observations, the \shk measurement failed due to a combination of poor seeing, scattered light, and overlapping orders at blue wavelengths.  These measurements are listed as N/A.}
\label{tb:RVdata_bai}
\end{deluxetable}

\begin{deluxetable}{lrrr}[h!]
\tablecaption{\ktwohbc Relative radial velocities, Keck-HIRES}
\tablehead{\colhead{BJD} & \colhead{RV} [\ms] & \colhead{Unc. [\ms]\tablenotemark{a}} & \colhead{\shk}}
\startdata
2457612.932644 & -5.04 & 1.89 & 0.149 \\
2457613.967264 & -3.25 & 1.58 & 0.147 \\
2457614.109833 & -2.35 & 1.50 & 0.150 \\
2457615.925879 & -5.08 & 1.71 & 0.148 \\
2457616.925922 & -3.58 & 1.66 & 0.150 \\
2457617.917421 & 4.13 & 1.53 & 0.148 \\
2457618.926340 & 5.95 & 1.53 & 0.147 \\
2457652.069904 & 10.72 & 1.53 & 0.143 \\
2457653.036506 & -0.86 & 1.64 & 0.139 \\
2457668.986188 & -11.42 & 1.72 & 0.137 \\
2457671.780051 & -24.84 & 1.94 & 0.150 \\
2457672.066034 & -5.93 & 1.66 & 0.152 \\
2457672.780348 & -6.86 & 1.69 & 0.153 \\
2457672.964502 & -12.31 & 1.61 & 0.153 \\
2457697.825599 & 5.88 & 1.77 & 0.142 \\
2457711.823439 & -16.01 & 2.31 & 0.150 \\
2457711.890113 & -4.35 & 1.52 & 0.139 \\
2457712.000267 & 1.74 & 1.94 & 0.132 \\
2457712.760228 & -0.49 & 1.75 & 0.137 \\
2457713.803918 & 4.51 & 1.85 & 0.140 \\
2457713.987377 & -7.84 & 1.55 & 0.141 \\
2457714.817690 & 1.62 & 1.57 & 0.136 \\
2457714.952333 & 8.26 & 2.07 & 0.134 \\
2457716.798647 & 1.70 & 1.91 & 0.139 \\
2457717.971107 & -7.35 & 2.40 & 0.124 \\
2457718.905031 & 4.66 & 2.21 & 0.132 \\
2457745.786321 & 1.55 & 1.80 & 0.144 \\
2457746.762857 & -5.93 & 1.74 & 0.135 \\
2457747.817094 & -9.04 & 1.83 & 0.134 \\
2457761.774749 & 6.95 & 1.63 & 0.138 \\
2457763.715781 & -0.96 & 1.57 & 0.142 \\
2457764.733619 & 2.21 & 1.79 & 0.139 \\
2457765.800317 & 5.55 & 2.94 & 0.111 \\
2457774.729422 & -0.41 & 1.53 & 0.138 \\
2457775.725113 & 2.53 & 1.65 & 0.138 
\enddata
\tablenotetext{a}{Uncertainties estimated from the dispersion in the radial velocity measured from 718 chunks. These uncertainties do not include ``jitter'' which is incorporated as a free parameter during the RV modeling ($\sigma_{\rm jit}$, Table \ref{tb:params_hbc}).} 
\label{tb:RVdata_hbc}
\end{deluxetable}

\section{Analysis}
\label{sec:analysis}
	
Here we describe the methods used to characterize planet host stars and to model our \ktwo light curves and RV time series.  Measured stellar parameters, light curve model parameters, and RV model parameters are listed in Tables \ref{tb:params_bai} and \ref{tb:params_hbc} for \ktwobai and \ktwohbc, respectively.

\subsection{Stellar characterization}
\label{sec:stellar_par}
From the iodine-free HIRES spectra, we measured the effective temperature (\Teff), surface gravity (\logg), and metallicity (\fe) of \ktwobai and \ktwohbc, using the updated ``Spectroscopy Made Easy'' (SME) analysis tool described in \citet{Brewer16}.  Previous comparison of SME results with astroseismic results demonstrated \logg values accurate to 0.05~dex \citep{Brewer15}.  Stellar masses and radii were estimated using the \texttt{isochrones} Python package \citep{Morton15a}, which fit our \Teff, \logg, and \fe measurements to a grid of models from the Dartmouth Stellar Evolution Database \citep{Dotter08}.  Posteriors were sampled using the \texttt{emcee} Markov Chain Monte Carlo (MCMC) package \citep{Foreman-Mackey13}.  The adopted uncertainties on stellar mass and radius correspond to 68.3\% (1$\sigma$) confidence intervals of the resulting posterior distributions.  For \ktwobai, we measure a mass \Mstar = \baistarmass\,\msun and radius \Rstar = \baistarrad\,\rsun.  These are consistent with the values \Mstar = 1.16 $\pm$ 0.05\,\msun, and \Rstar = 1.71 $\pm$ 0.14\,\rsun reported by \citet{Crossfield16}, who used the SpecMatch algorithm \citep{Petigura15b} instead of SME.  For \ktwohbc, we measure a mass of \hbcstarmass\,\msun and radius of \hbcstarrad\,\rsun.  \citet{Adams16} measured \Mstar = 0.93 $\pm$ 0.01 \msun, which is consistent with our measurement, but they estimated \Rstar = 0.83 $\pm$ 0.04\,\rsun, which is smaller than our measurement at the $\sim$ 2.5-$\sigma$ level (see discussion in \S\ref{sec:composition_K2106}).

To test for spectroscopic blends, we used the algorithm of \citet{Kolbl15} to search for multiple sets of stellar lines.  For both  \ktwobai and \ktwohbc, we ruled out the possibility of companions in the $0\farcs87 \times 14\arcsec$ HIRES slit with \Teff = 3400--6100\,K, down to 1\% contrast in V and R bands, and $\Delta{RV}$ $> 10$ \kms.

The magnetic activity of each star was assessed by measuring \shk indices using the Ca II H \& K spectral lines \citep{Isaacson10}.  The \shk measurements are listed in Tables \ref{tb:RVdata_bai} and \ref{tb:RVdata_hbc} for \ktwobai and \ktwohbc respectively. The median \shk values from all spectra are \baistarshk and \hbcstarshk.  The measured \Teff and \shk were converted into \lrphk values, a metric of the Ca II flux relative to the photospheric continuum \citep{Middelkoop82, Noyes84}.   We measure median \lrphk values of \baistarrphk and \hbcstarrphk dex, consistent with magnetically quiet stars from the California Planet Search \citep{Isaacson10}.  For comparison, the Sun ranges from \lrphk = $-5.05$ dex to $-4.85$ dex over a typical magnetic cycle \citep{Meunier10}.  

\begin{figure}
        \subfigure[]{%
            \label{fig:AO_image}
            \includegraphics[width=1.0\columnwidth]{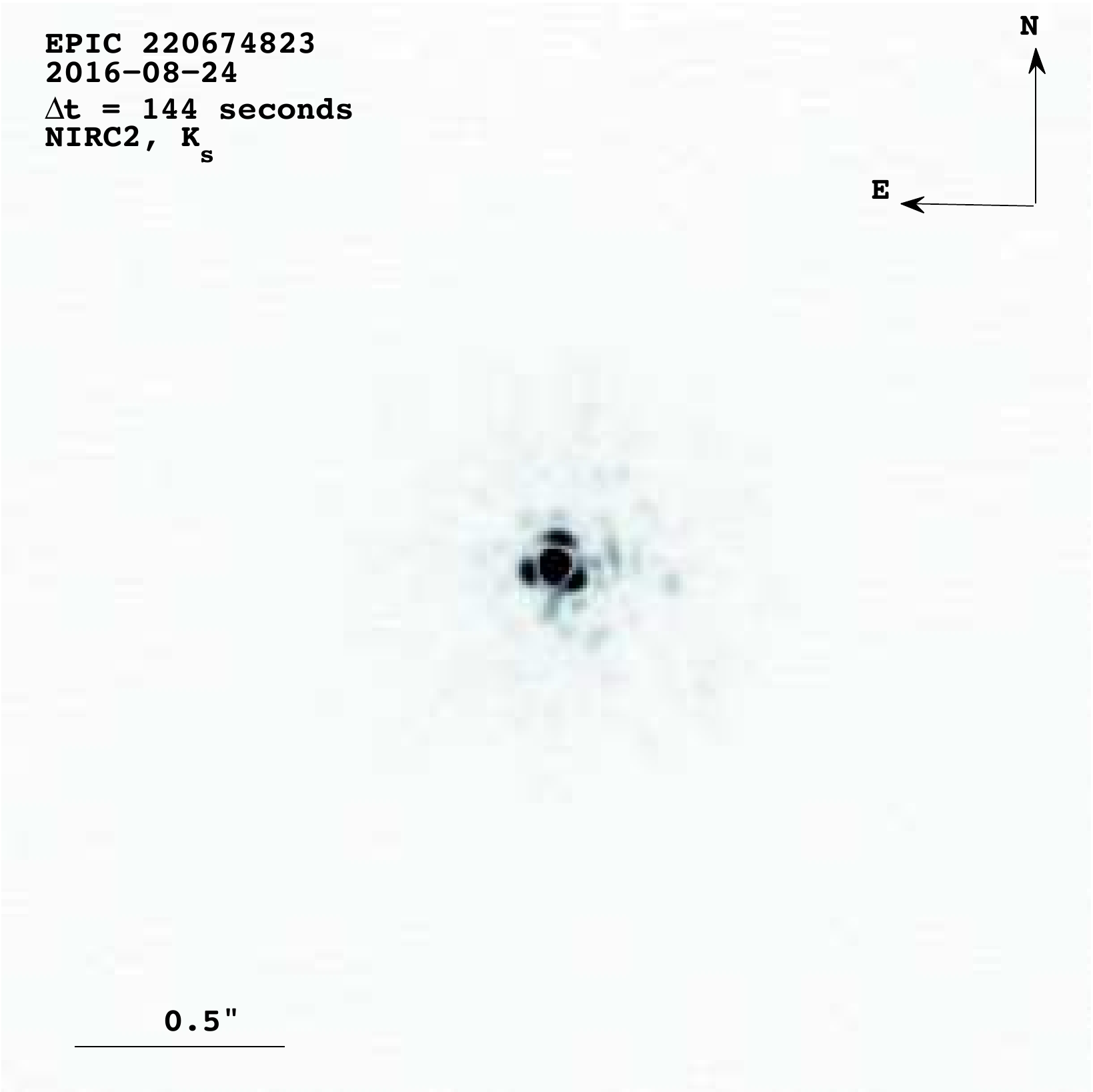}
        }\\%
        \subfigure[]{%
           \label{fig:AO_contrast}
           \includegraphics[width=1.0\columnwidth]{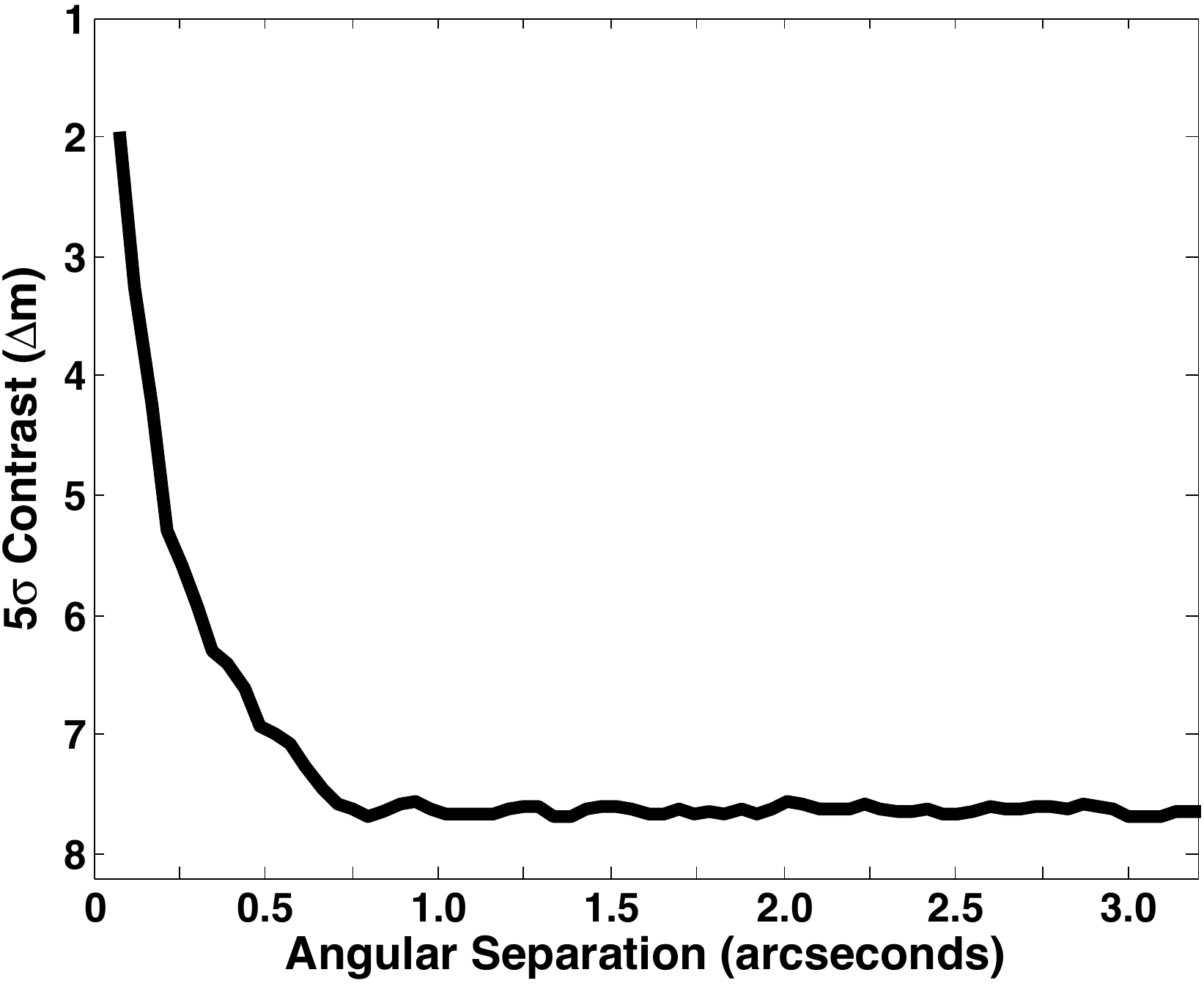}
         }
    \caption{Keck/NIRC2 K$_s$-band adaptive optics imaging of \ktwohbc. \textit{(a)} Reduced image, showing no evidence of secondary stars. \textit{(b)} 5$\sigma$ contrast limits. 
     }%
   \label{fig:AO}
\end{figure}

Our NIRC2 images were processed using a standard flat-field, background subtraction, and image stacking techniques \citep[e.g][]{Crepp12}. Figure \ref{fig:AO_image} displays the final reduced image and angular scale. Both raw and stacked images were examined for companion sources. A speckle to the right of the host star was ruled out as a companion as stacked images in the J-band filter showed it moving as a function of wavelength. Figure \ref{fig:AO_contrast} shows the sensitivity to nearby companions. Contrast levels reach $\Delta$K = 7.7 for separations beyond $0\farcs75$.  \citet{Adams16} achieve similar contrast limits from K-band observations of \ktwohbc, also with Keck/NIRC2 AO.

\subsection{Light curve analysis}
\begin{figure*}
    \centering
    \includegraphics[width=2.0\columnwidth]{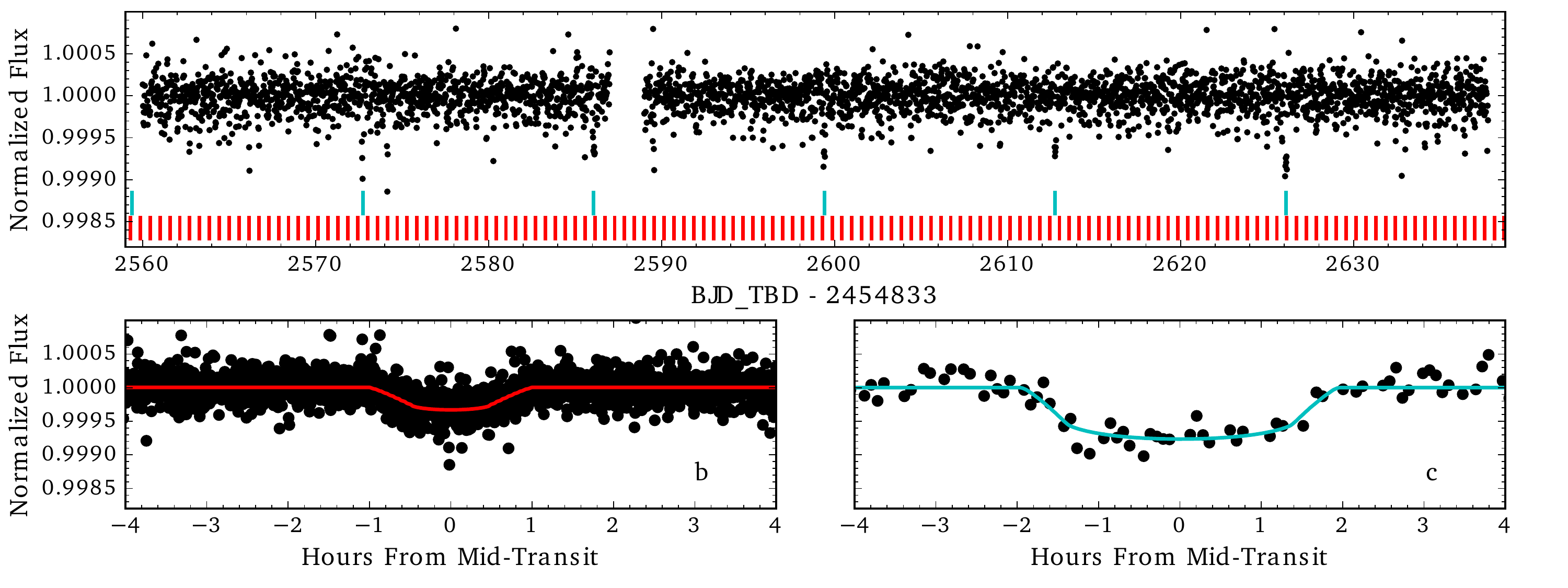}
    \caption{\textit{Top}: Calibrated \ktwo photometry for \ktwohbc. Vertical ticks indicate the locations of each planets' transits.  Bottom: Phase-folded photometry and best-fit light curves for each of the two planets.}
    \label{fig:LC_hbc}
\end{figure*}

We fit transit models to the detrended \ktwohbc light curve using the same MCMC analysis described in \citet{Crossfield16}.  In brief, our code employs the Markov Chain Monte Carlo (MCMC) package \texttt{emcee} \citep{Foreman-Mackey13} and model light curves are generated using the Python package \texttt{BATMAN} \citep{Kreidberg15}.  The model parameters are:  time of conjunction ($T_{\rm{conj}}$ ), orbital period, eccentricity, inclination, and longitude of periastron ($P$ $e$, $i$, and $\omega$), scaled semimajor axis (\ars), ratio of planet radius to stellar radius (\rprs), a single multiplicative offset for the absolute flux level, and quadratic limb-darkening coefficients ($u_0$ and $u_1$).  The detrended \ktwohbc light curve and fitted transit models for planets b and c are shown in Figure \ref{fig:LC_hbc} 
  
\subsection{RV Analysis}

\subsubsection{Methodology}

To analyze the RV time-series of \ktwobai and \ktwohbc, we used the RV fitting package \texttt{RadVel} (B. Fulton \& E. Petigura, in prep.), which is publicly available on GitHub\footnote{https://github.com/California-Planet-Search/radvel \\ http://radvel.readthedocs.io/en/master/index.html}.  \texttt{RadVel} is written in object-oriented Python.  It uses a fast Kepler equation solver written in C and the affine-invariant sampler \citep{Goodman10} of the \texttt{emcee} package \citep{Foreman-Mackey13}.  \texttt{RadVel} is easily adaptable to a variety of maximum-likelihood fitting and MCMC applications. The standard version allows for modeling of multi-planet, multi-instrument RV time-series, and assumes no interaction between planets \citep[e.g.][]{Sinukoff17a}.


We adopt the same likelihood function for RV modeling as \citet{Howard14}:
\begin{align}
    \begin{split}
        \ln{\mathcal{L}} ={} -\sum_{i} & \left[\frac{\left(v_i-v_m(t_i)\right)^2}{2\left(\sigma_i^2+\sigma_{\mathrm{jit}}^2\right)}\right. \\
        {} & + \ln{\sqrt{2\pi\left(\sigma_i^2 + \sigma_{\mathrm{jit}}^2\right)}}\left.\vphantom{\frac{\left(v\right)^2}{\left(v\right)}}\right],
    \end{split} 
\end{align}
where $v_i$ and $\sigma_i$ are the \textit{i}th RV measurement and corresponding uncertainty, and $v_m$($t_i$) is the Keplerian model velocity at time $t_i$.  The same RV model parameters are used as MCMC step parameters.  Before starting the MCMC exploration, we first use the minimization technique of \citet{Powell64} to find the maximum-likelihood model.  Fifty parallel MCMC chains (``walkers'') are then initialized by perturbing each of the free parameters from the maximum likelihood values by as much as 3\%.   An initial round of MCMC exploration continues until the Gelman-Rubin (GR) statistic \citep{Gelman92} drops below 1.10, at which point the chains are reset.  Following this burn-in phase, the remaining chains are kept and the MCMC run proceeds until the GR $<$ 1.03 and the $T_{z}$ statistic \citep{Ford06} exceeds 1000 for all free parameters.  This ensures that the chains are well-mixed and converged. 

The adopted basis for our RV model for both \ktwobai and \ktwohbc is: \{$P$, $T_{\rm{conj}}$ , $K$, $\gamma$\}, where $P$ is orbital period, $T_{\rm{conj}}$  is the time of conjunction, $K$ is the RV semi-amplitude and $\gamma$ is a constant RV offset.  For \ktwohbc, we fit for $P$, $T_{\rm{conj}}$, and $K$ of both planets.  We lock the orbital periods and phases at the photometrically measured values in Tables \ref{tb:params_bai} and \ref{tb:params_hbc}.  Since the orbital ephemeris is tightly constrained from photometry, it made no difference whether we fixed the ephemeris or assigned Gaussian priors according to uncertainties on $P$ and $T_{\rm{conj}}$.  When testing non-circular orbits, we include two additional model parameters, $\sqrt{e}\cos{\omega}$ and $\sqrt{e}\sin{\omega}$, where $e$ is the orbital eccentricity and $\omega$ is the longitude of periapsis of the star's orbit.  This parameterization mitigates the Lucy-Sweeney bias toward non-zero eccentricity \citep{Lucy71, Eastman13}.  We also search for additional bodies at orbital periods beyond the duration of RV observations by testing RV models that include a constant acceleration term, $\mathrm{d}v/\mathrm{d}t$ (i.e. a linear trend in the RV time series).  To assess whether the addition of eccentricity and constant acceleration parameters are warranted, we use the Bayesian Information Criterion (BIC). When comparing models, we lock the RV jitter at the values in Tables \ref{tb:params_bai} and \ref{tb:params_hbc}.

In \S\ref{sec:planetsearch}, we discuss our search for additional planets in these two systems.  We found no conclusive evidence for additional planets.

\subsubsection{\ktwobai}

\begin{figure*}
    \centering
    \includegraphics[width=1.7\columnwidth]{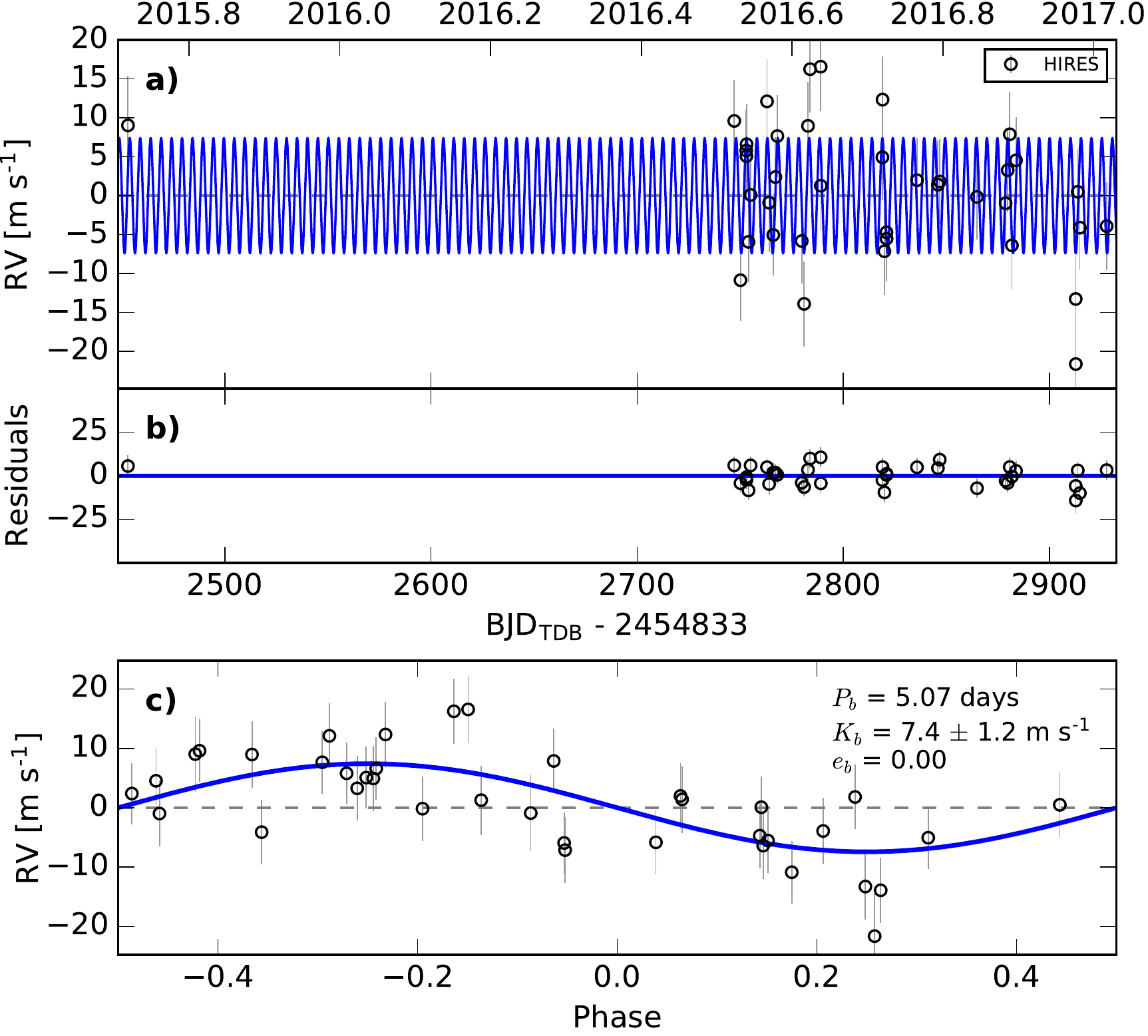}
    \caption{Single-planet RV model of \ktwobai, assuming a circular orbit and adopting the ephemeris from transit fits.  \textit{a)} The RV time-series. Open black circles indicate Keck/HIRES data.  The solid blue line corresponds to the most likely model. Note that the orbital parameters listed in Table \ref{tb:params_bai} are the median values of the posterior distributions. Error bars for each independent dataset include an RV jitter term listed in Table \ref{tb:params_bai}, which are added in quadrature to the measurement uncertainties. \textit{b)} Residuals to the maximum-likelihood fit. \textit{c)} The RV time-series phase folded at the orbital period of \ktwobai{b}.}
    \label{fig:RVfit_bai}
\end{figure*}

After testing several different RV model parameterizations for \ktwobai, we adopt a circular orbit (sinusoidal) model with zero acceleration ($\mathrm{d}v/\mathrm{d}t\equiv 0$). The adopted RV parameters for \ktwobai are listed in Table \ref{tb:params_bai}, including $K=$\baibk\,\ms.  The maximum likelihood RV fit is shown in Figure \ref{fig:RVfit_bai}.  When the orbital eccentricity is allowed to float, the MCMC fit yields $e$=0.10$^{+0.13}_{-0.07}$, and a planet mass consistent with the circular orbit model.  The change in the BIC is $\Delta${}BIC = BIC$_{\mathrm{ecc}}-$BIC$_{\mathrm{circ}}$ = 1.0, which indicates that the fit does not improve enough to justify the additional free parameters \citep{Kass95}.   Similarly, introducing $\mathrm{d}v/\mathrm{d}t$ as a free parameter yields $\Delta${}BIC = BIC$_{\mathrm{d}v/\mathrm{d}t}-$BIC$_{\mathrm{d}v/\mathrm{d}t\equiv 0}$ = $-$0.7, indicating no preference for the more complex model.  Each of the different RV models that were tested resulted in a planet mass within 0.5$\sigma$ of the adopted value.  

\subsubsection{\ktwohbc}
\begin{figure*}
    \centering
    \includegraphics[width=1.7\columnwidth]{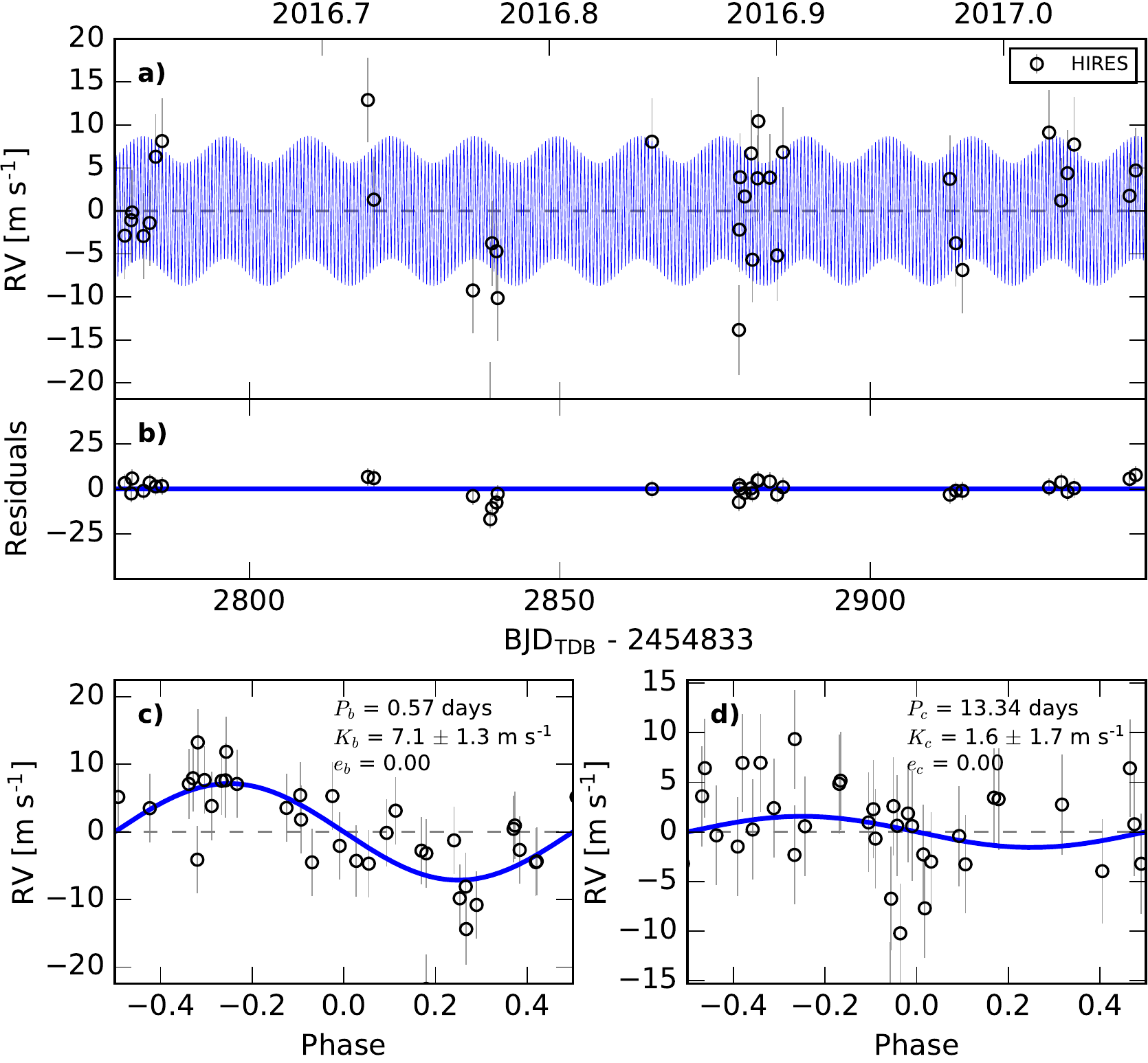}
    \caption{Two-planet RV model of \ktwohbc, assuming circular orbits and adopting the ephemerides from transit fits.  Details are same as Figure \ref{fig:RVfit_bai}, with panels c and d showing the phase-folded light curves for planets b and c, after subtracting the signal of the other planet.  We do not make a statistically significant measurement of the mass of planet c.}
   \label{fig:RVfit_hbc}
\end{figure*}

The adopted RV model for \ktwohbc is the sum of two sinusoids (two circular orbits), with $\mathrm{d}v/\mathrm{d}t \equiv 0$.  The fitted RV parameters for \ktwohbc are listed in Table \ref{tb:params_hbc} and the adopted RV fit is displayed in Figure \ref{fig:RVfit_hbc}. Overall, the choice of model did not significantly affect the planet mass measurements --- all of the RV models yielded planet mass constraints consistent with the adopted values. For planet b, we measure $K$ = \hbcbk\,\ms, for a $5.5\sigma$ detection.  For planet c, we measure $K$ = \hbcck\,\ms, which is not a reliable detection.  From the posterior distribution, we place an upper limit, $K~<~$\hbccthreesigk\ms ($\Mp ~<~$\hbccthreesigmp\,\mearth) at 99.7\% confidence.  Due to its proximity to the host star, the orbit of \ktwohbc{b} has likely been circularized by tidal interactions with the star:  We compute a circularization timescale of $\approx$ 6000 years using \citep{Goldreich66} assuming the same a tidal quality factor $Q$ = 100 estimated for terrestrial planets in the Solar System \citep{Goldreich66, Henning09, Lainey16}.  Nevertheless, we tested a fit to the RV time series in which the eccentricity of planet b was allowed to float.  The MCMC fit yielded $e$ = \ensuremath{0.11^{+0.11}_{-0.08}}, and a planet mass consistent with the best circular orbit model. Moreover, the eccentric model is not statistically favored ($\Delta${}BIC = 0.1).  When the eccentricity of planet c was allowed to float, the preferred eccentricity was 0.75 and the MCMC chains did not converge. Any orbit $e$ $\gtrsim$ 0.35 would cross the stellar surface.    We also ran a trial with $\mathrm{d}v/\mathrm{d}t$ as a free parameter, but found this additional model complexity was not statistically warranted ($\Delta${}BIC = 0.2).  Finally, since planet c was not significantly detected, we also tried fitting for planet b alone but the measured mass changes by $<$ 0.5$\sigma$.    

There are several possible reasons why we do not detect the RV signal of planet c.  One possibility is that $K$ is sufficiently small that more data are needed to securely detect the planet.  Alternatively, stellar activity on the timescale of the planet's orbital period (13 days) could partially wash out the planet signal. However, our \lrphk measurement of \hbcstarrphk indicates a magnetically quiet star.  Finally, the star might host additional planets not included in our RV model.  

\subsubsection{Search for Additional Planets}
\label{sec:planetsearch}

\begin{figure*}
    \centering
    \subfigure[]{
    \includegraphics[width=1.0\columnwidth]{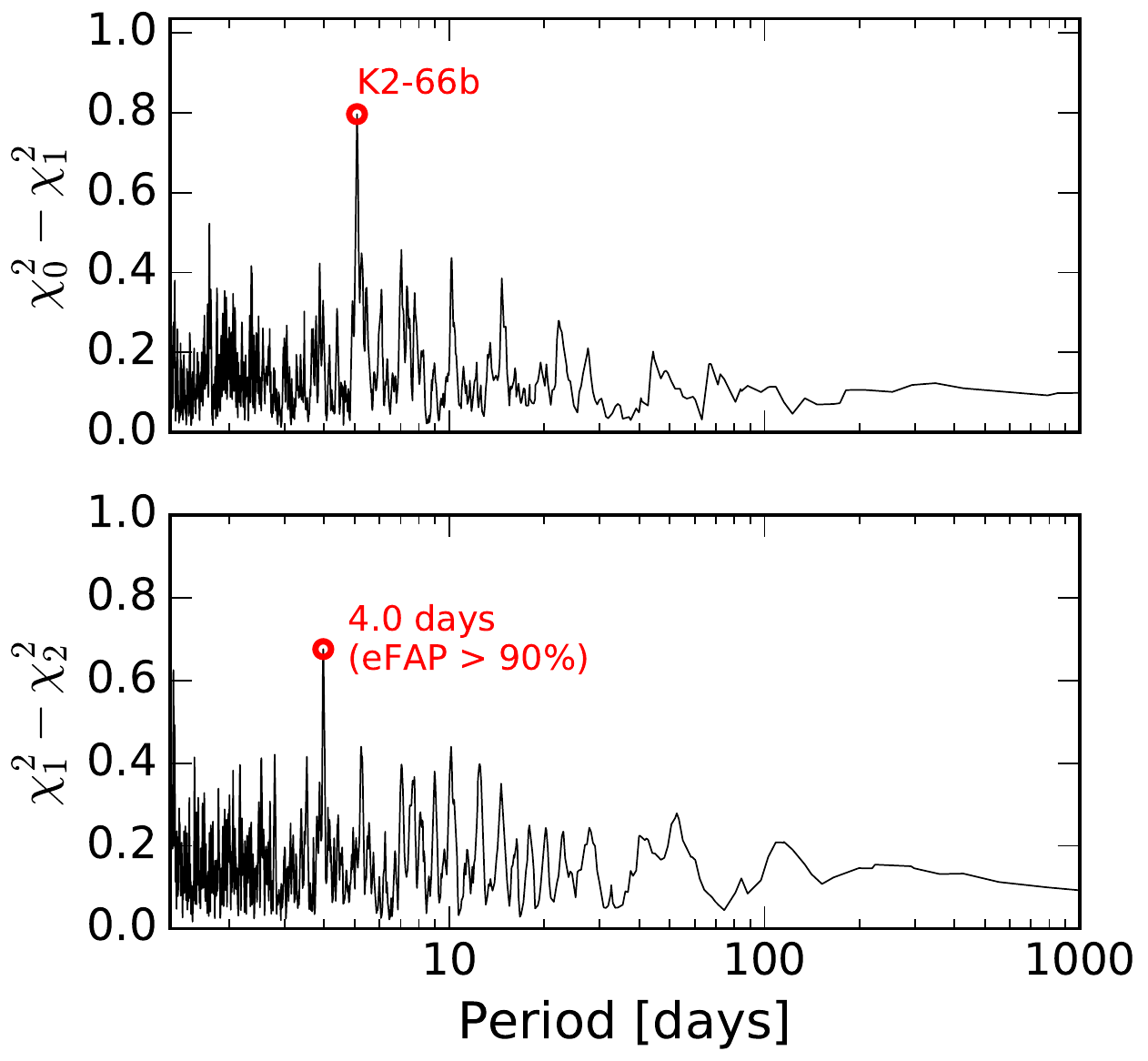}
    }
     \subfigure[]{
    \includegraphics[width=1.0\columnwidth]{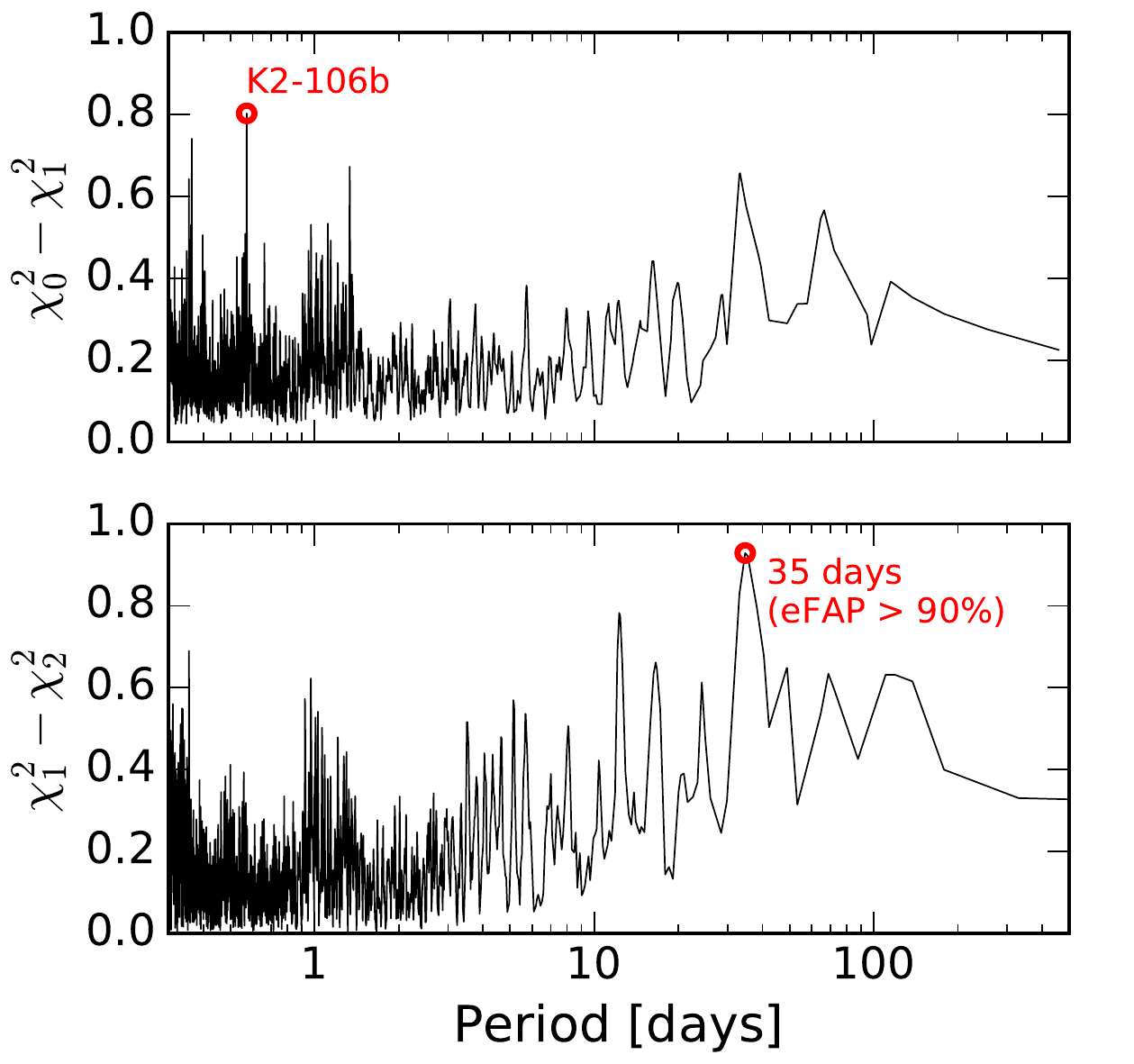}
    }
    \caption{Two-dimensional Keplerian Lomb-Scargle periodograms of the measured RV time series of \textbf{a)} \ktwobai and \textbf{b)} \ktwohbc.  Values on the vertical axis represent the difference in $\chi^2$ between an $N$-planet model ($\chi^2_N$ ) and an $N$+1 planet model ($\chi^2_{N+1}$) at each period.  The tallest peaks in the $N$ = 0 cases (top panels) correspond to the periods of known transiting planets, as labeled. For the $N$ = 1 cases (bottom panels), empirical false alarm probabilities (eFAPs) for the tallest peaks are $>$ 90\%.  They are likely to be spurious signals rather than the signals of additional planets.}
   \label{fig:pergrams}
\end{figure*}

We conducted a search for additional planets in both systems using the planet search algorithm described in \citet{Howard16}, which utilizes a two-dimensional Keplerian Lomb-Scargle periodogram \citep[2DKLS,][]{Otoole09}. The periodogram values represent the difference in $\chi^2$ between an $N$-planet model ($\chi^2_N$ ) and an $N$+1 planet model ($\chi^2_{N+1}$) for each orbital period value. When searching for the first planet in a given system we compare $\chi^2$ for a 1-planet model to $\chi^2$ for a flat line.  Figure \ref{fig:pergrams} shows the periodograms for $N$ = 0 and $N$ = 1. We estimate an empirical false alarm probability (eFAP) for any peaks in the 2DKLS periodogram by fitting a log-linear function to a histogram of periodogram values. 

For \ktwobai, we find no evidence of additional planet signals in the RV time series.  In the $N$ = 0 case, the tallest peak in the periodogram occurs at 5.1 days, corresponding to the known transiting planet \ktwobai{b}.  For $N$ = 1, which tests the 2-planet hypothesis, the tallest peak is at $P$ = 4.0 days and has eFAP $>$ 90\%.  We note that when we tested a 2-planet RV model with an initial period guess of 4.0 days for the second Keplerian, the measured RV semi-amplitude for K2-66b remains consistent with the adopted 1-planet model at $\approx$ 0.3 $\sigma$.  Therefore, even if there is an additional planet at $P$ $\approx$ 4 days, it does not significantly influence our mass measurement for \ktwobai{b}.

Similarly, for \ktwohbc, our search for additional planets in the RV time-series yields a null result.  The periodogram for $N$ = 0 has a global maximum at the orbital period of \ktwohbc{b} (0.57 days).  The $N$ = 1 periodogram does not have any significant peaks --- the tallest is at $P$ = 35 days with eFAP $>$ 90\%.  We conclude that more RV data are needed to confidently detect any additional bodies orbiting \ktwohbc.  We note that the measured RV semi-amplitude for \ktwohbc{b} changes by $<$ 0.5$\sigma$ when a 3-planet RV model is tested with an initial period guess of 35 days for the third Keplerian.  Thus, even if there is an additional planet at $P$ $\approx$ 35 days, it has a negligible effect on our mass measurement for \ktwohbc{b}.


\section{Results \& Discussion}
\label{sec:discussion}

\subsection{No Significant Dilution}
\label{sec:dilution}

\begin{figure}
        \subfigure[]{%
            \label{fig:companion_constraint_bai}
            \includegraphics[width=1.0\columnwidth]{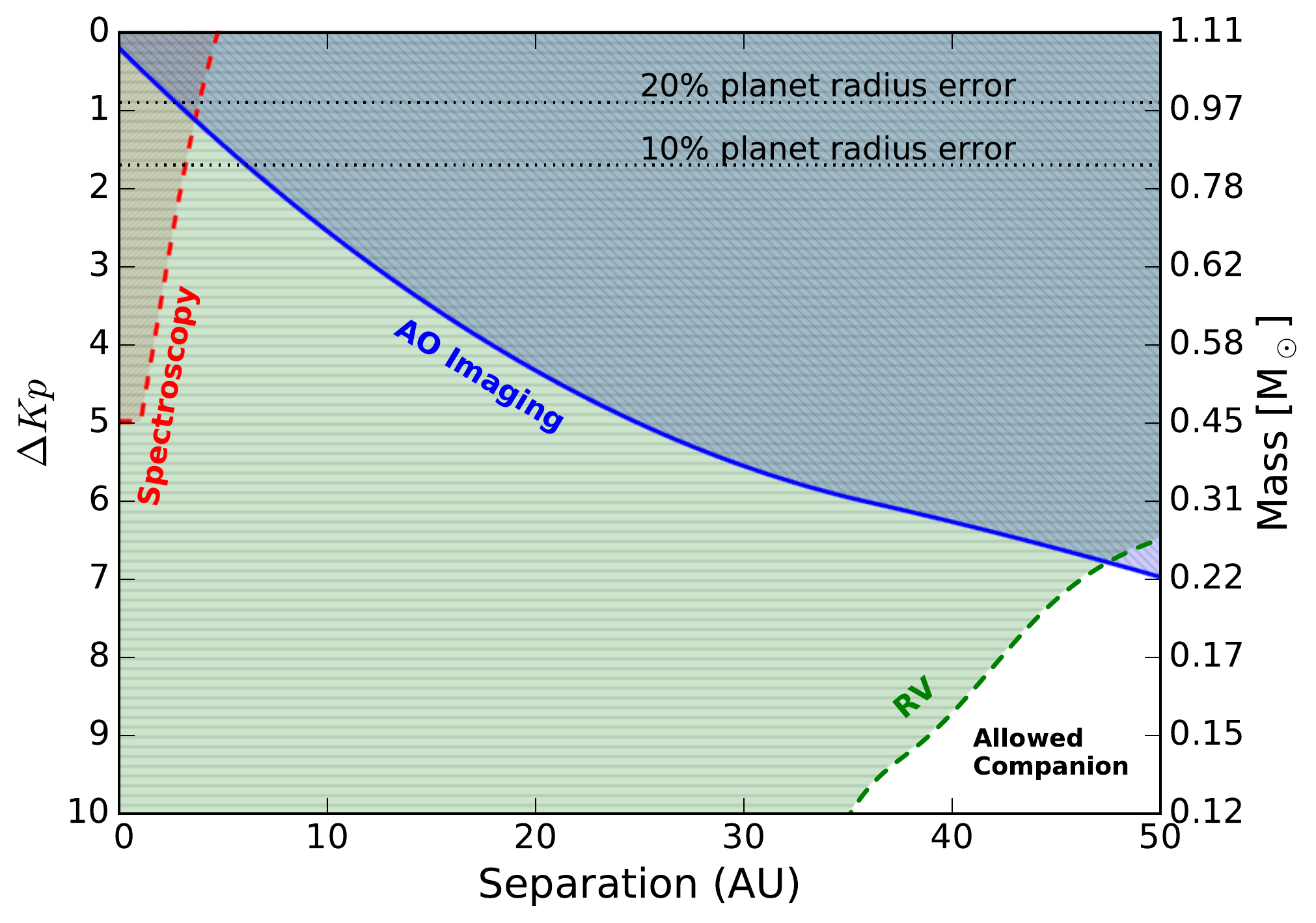}
        }\\%
        \subfigure[]{%
           \label{fig:companion_constraint_hbc}
           \includegraphics[width=1.0\columnwidth]{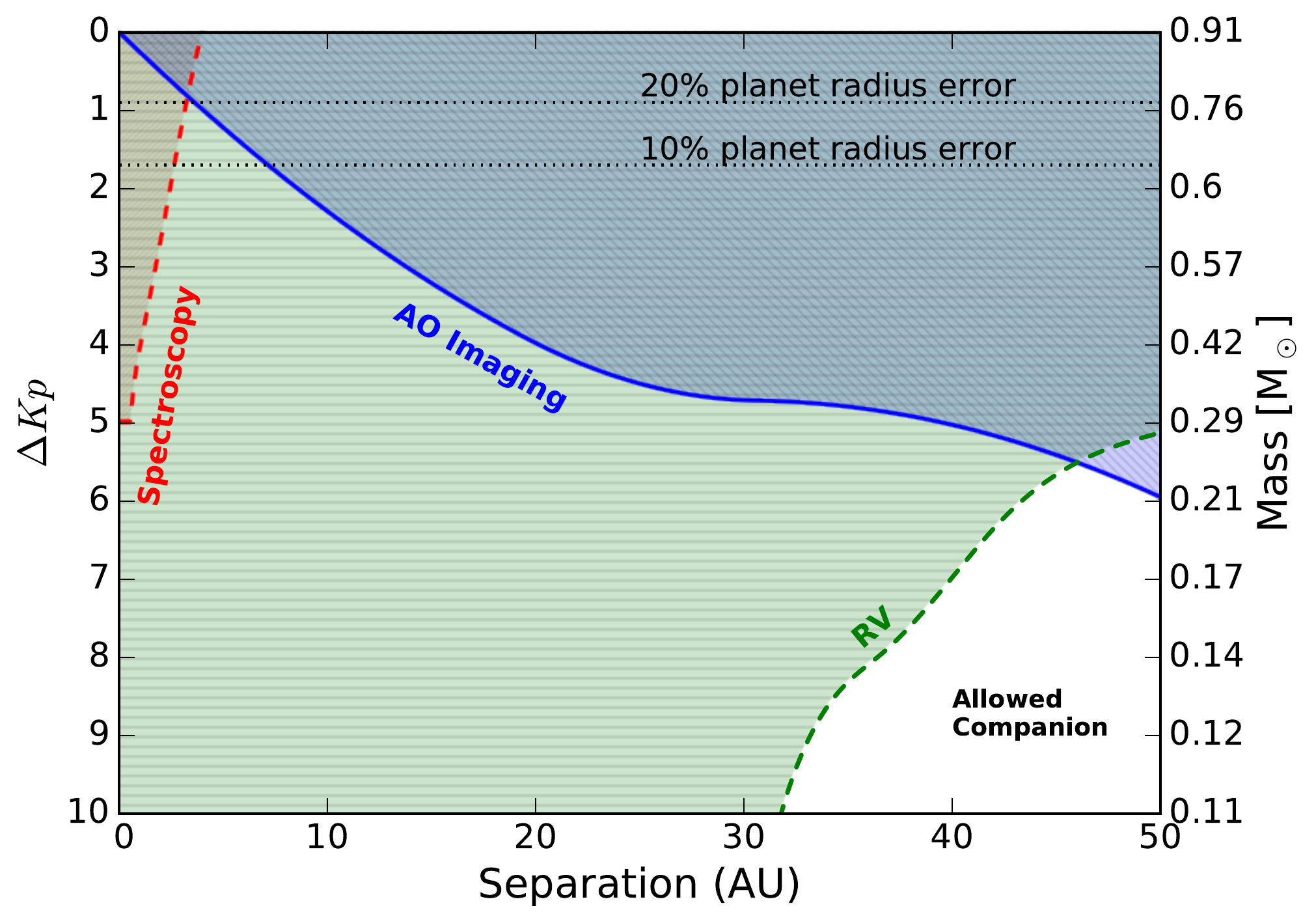}
         }
    \caption{Constraints on the presence of other stars in the photometric aperture for \textit{(a)} \ktwobai and \textit{(b)} \ktwohbc, which would dilute the measure transit depth.  The vertical axes show companion brightness contrast and companion mass plotted against orbital separation. NIRC2 AO imaging excludes companions in the hatched blue region, assuming distances of 400 pc and 250 pc to \ktwobai and \ktwohbc, respectively. The dashed red line shows the limits of our search for secondary lines in the HIRES spectrum. Companions in the hatched green region would induce a linear RV trend larger than the 3-$\sigma$ upper limit determined from the RV time-series, assuming a circular, edge-on orbit. The horizontal dotted lines represent companion contrasts at which the dilution of the observed transit depths of \ktwobai{b} and \ktwohbc{b} would cause planet radii to be overestimated by 10\% and 20\%.  Together, AO imaging and spectroscopy, and RVs rule out companions that would cause systematic errors of $>$ 10\% in planet radius with high confidence (see \S \ref{sec:dilution} for discussion)
     }%
   \label{fig:companion_constraint}
\end{figure}

Our RV detections of \ktwobai{b} and \ktwohbc{b} confirm that they are bonafide planets.  To verify that the planet radius measurements are accurate, we investigated the possibility that the photometric aperture contains a blend of multiple stars.  Blends would dilute the transit depth, causing the planet radius to be underestimated \citep{Ciardi15}.  Figure \ref{fig:companion_constraint} shows blend constraints from the spectroscopic analysis, AO images, and RV measurements.  Together, these rule out the presence of companions that would significantly alter the measured planet radii.   Contrasts in the NIRC2-AO bandpass were converted to the Kepler bandpass and to companion masses using $riJHK$ photometric calibrations of \citet{Kraus07}.  A blend with Kepler-band contrast \dkp $\lesssim$ 2 mag is required for a 10\% error in the measured planet radius.   Such companions within $\sim$ 100 AU of \ktwobai or \ktwohbc would have been detected as a linear trend in the RV time-series and would have been detected inside $\sim$ 5 AU as secondary lines in the HIRES spectrum.  AO imaging rules out problematic companions beyond $\sim$ 10 AU.  We note that the plotted constraints from RV observations use Equation 1 of \citet{Winn10}, and conservatively assume \dvdt values equal to the 3-$\sigma$ upper limits obtained when \dvdt is included as a free model parameter.  The only conceivable problematic blend that would be undetected is a companion near apastron of a highly eccentric orbit (hence low \dvdt), at an orbital phase of low projected separation (hence undetected in AO images) and with a spectrum similar to that of the primary star (hence undetected spectral lines).  However, such a scenario is highly improbable and we conclude that the likelihood of a problematic blend is negligibly low.

\subsection{Planetary Bulk Compositions}

\begin{deluxetable}{lrl}
\tablecaption{\ktwobai system parameters}
\tablehead{\colhead{Parameter} & \colhead{Value} & \colhead{Units}}
\startdata
\hline
\sidehead{{\bf Stellar Parameters}} 
$V$ & \baistarvmag & mag \\
\Teff  &   \baistarteff    & K      \\
\logg  &   \baistarlogg    & dex      \\
\fe    &   \baistarfe      & dex      \\ 
\vsini &   \baistarvsini   & \kms         \\
\Mstar &   \baistarmass    & \msun    \\
\Rstar &   \baistarrad     & \rsun    \\
\hline 
\sidehead{{\bf Planet b} }
\sidehead{{Transit Model} }
$P$          & \baibper   & days               \\
$T_{\rm{conj}}$  & \baibtt    & BJD              \\
\rprs        & \baibrprs  & ---                \\
\rsa         & \baibrsa   & ---          \\
\uzero   &   \baibuzero & ---          \\
\uone   &   \baibuone & ---          \\
$b$       &    \baibb     & ---         \\
$i$  &   \baibincdeg & deg          \\
\tdur & \baibtdur & hrs          \\
\rhostar & \baibrhostar & \gcc  \\
\sidehead{{RV Model (circular orbit assumed)} } 
$K$          & \baibk     & \ms             \\
\sidehead{{Derived Planet Parameters} }
$a$ & \baiba & au  \\
$S_{\rm{inc}}$ & \baibfinc  & $S_\oplus$      \\
$T_{\rm{eq}}$ & \baibteq  & K      \\
\Rp          & \baibrp    & \rearth         \\
\Mp          & \baibmp    & \mearth         \\
\rhop        & \baibrhop  & \gcc            \\
\hline
\sidehead{\bf{Other}}
$\gamma$        & \baibgammaj      & \ms   \\
$\sigma_{\rm jit}$        & \baibjitj        & \ms   \\
\enddata
\tablecomments{$S_{\rm{inc}}$ = incident flux, $T_{\rm{conj}}$ = time of conjunction. $T_{\rm{eq}}$ = equilibrium temperature, assuming albedo = 0.3}
\label{tb:params_bai}
\end{deluxetable}

\begin{deluxetable}{lrlr}
\tablecaption{\ktwohbc system parameters}
\tablehead{\colhead{Parameter} & \colhead{Value} & \colhead{Units}}
\startdata
\hline
\sidehead{{\bf Stellar Parameters}} 
$V$ & \hbcstarvmag & mag \\
\Teff  &   \hbcstarteff    & K      \\
\logg  &   \hbcstarlogg    & dex      \\
\fe    &   \hbcstarfe      & dex      \\ 
\vsini &   \hbcstarvsini   & \kms         \\
\Mstar &   \hbcstarmass    & \msun    \\
\Rstar &   \hbcstarrad     & \rsun    \\
\hline 
\sidehead{{\bf Planet b} }
\sidehead{{Transit Model} }
$P$          & \hbcbper   & days               \\
$T_{\rm{conj}}$  & \hbcbtt    & BJD              \\
\rprs        & \hbcbrprs  & ---                \\
\rsa         & \hbcbrsa   & ---          \\
\uzero   &   \hbcbuzero & ---          \\
\uone   &   \hbcbuone & ---          \\
$b$       &    \hbcbb     & ---         \\
$i$  &   \hbcbincdeg & deg          \\
\tdur & \hbcbtdur & hrs          \\
\rhostar & \hbcbrhostar & \gcc  \\
\sidehead{{RV Model (circular orbit assumed)} } 
$K$          & \hbcbk     & \ms             \\
\sidehead{{Derived Planet Parameters} }
$a$ & \hbcba & au  \\
$S_{\rm{inc}}$ & \hbcbfinc  & $S_\oplus$      \\
$T_{\rm{eq}}$ & \hbcbteq  & K      \\
\Rp          & \hbcbrp    & \rearth         \\
\Mp          & \hbcbmp    & \mearth         \\
\rhop        & \hbcbrhop  & \gcc            \\
\hline
\sidehead{\bf Planet c}
\sidehead{{Transit Model} }
$P$          & \hbccper   & days               \\
$T_{\rm{conj}}$  & \hbcctt    & BJD              \\
\rprs        & \hbccrprs  & ---                \\
\rsa         & \hbccrsa   & ---          \\
\uzero   &   \hbccuzero & ---          \\
\uone   &   \hbccuone & ---          \\
$b$       &    \hbccb     & ---         \\
$i$  &   \hbccincdeg & deg          \\
\tdur & \hbcctdur & hrs          \\
\rhostar & \hbccrhostar & \gcc  \\
\sidehead{{RV Model (circular orbit assumed)} }
$K$          & \hbcck     & \ms             \\
\sidehead{{Derived Planet Parameters} }
$a$ & \hbcca & au  \\
$S_{\rm{inc}}$ & \hbccfinc  & $S_\oplus$      \\
$T_{\rm{eq}}$ & \hbccteq  & K      \\
\Rp          & \hbccrp    & \rearth         \\
\Mp          & \hbccmp    & \mearth         \\
\rhop        & \hbccrhop  & \gcc            \\
\hline
\sidehead{\bf{Other}}
$\gamma$        & \hbcbgammaj      & \ms   \\
$\sigma_{\rm jit}$        & \hbcbjitj        & \ms  \\
\enddata
\tablecomments{$S_{\rm{inc}}$ = incident flux, $T_{\rm{conj}}$ = time of conjunction. $T_{\rm{eq}}$ = equilibrium temperature, assuming albedo = 0.3}
\label{tb:params_hbc}
\end{deluxetable}


\begin{figure}
        \subfigure[]{%
            \label{fig:MR_all}
            \includegraphics[width=1.0\columnwidth]{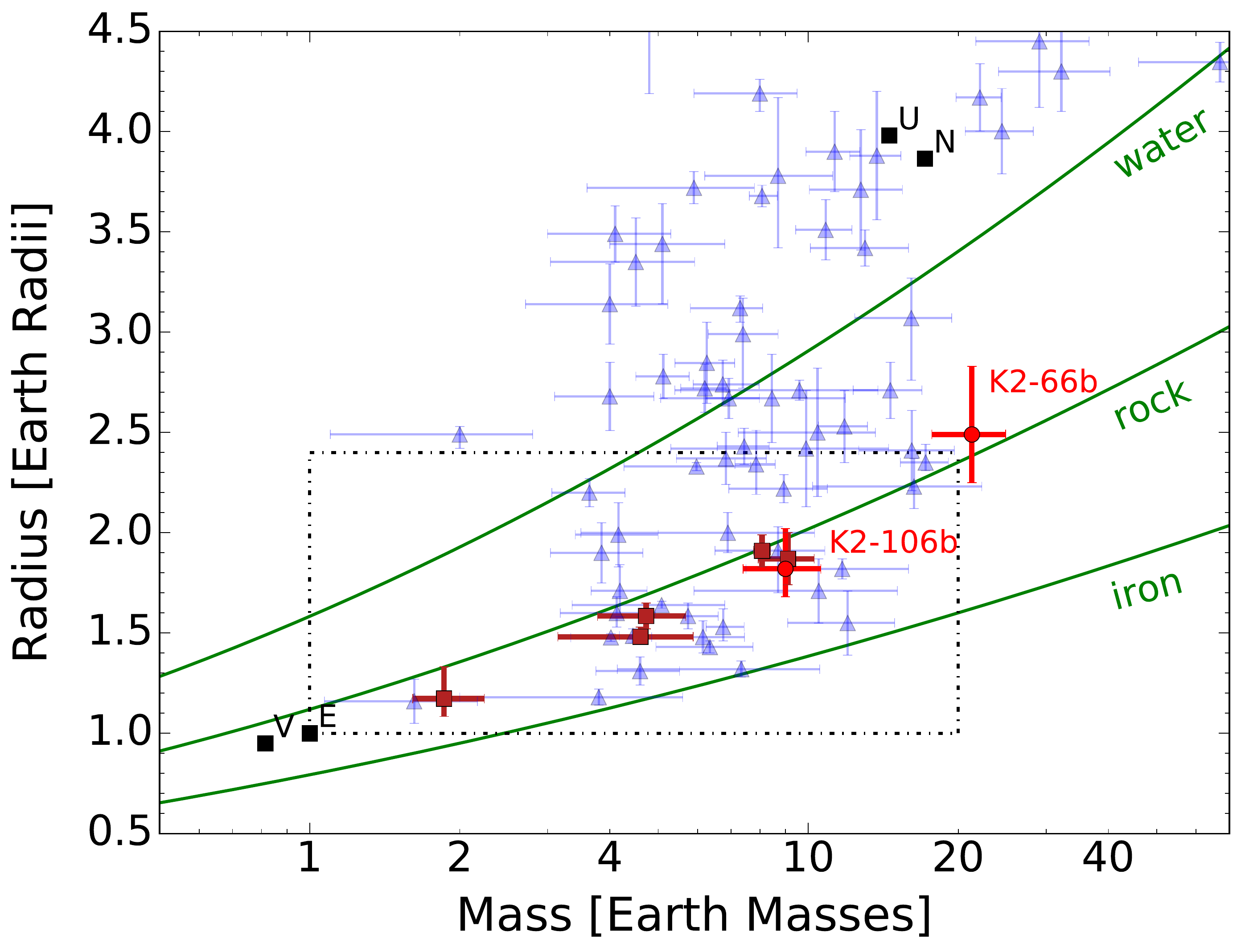}
        }\\%
        \subfigure[]{%
           \label{fig:MR_USP}
           \includegraphics[width=1.0\columnwidth]{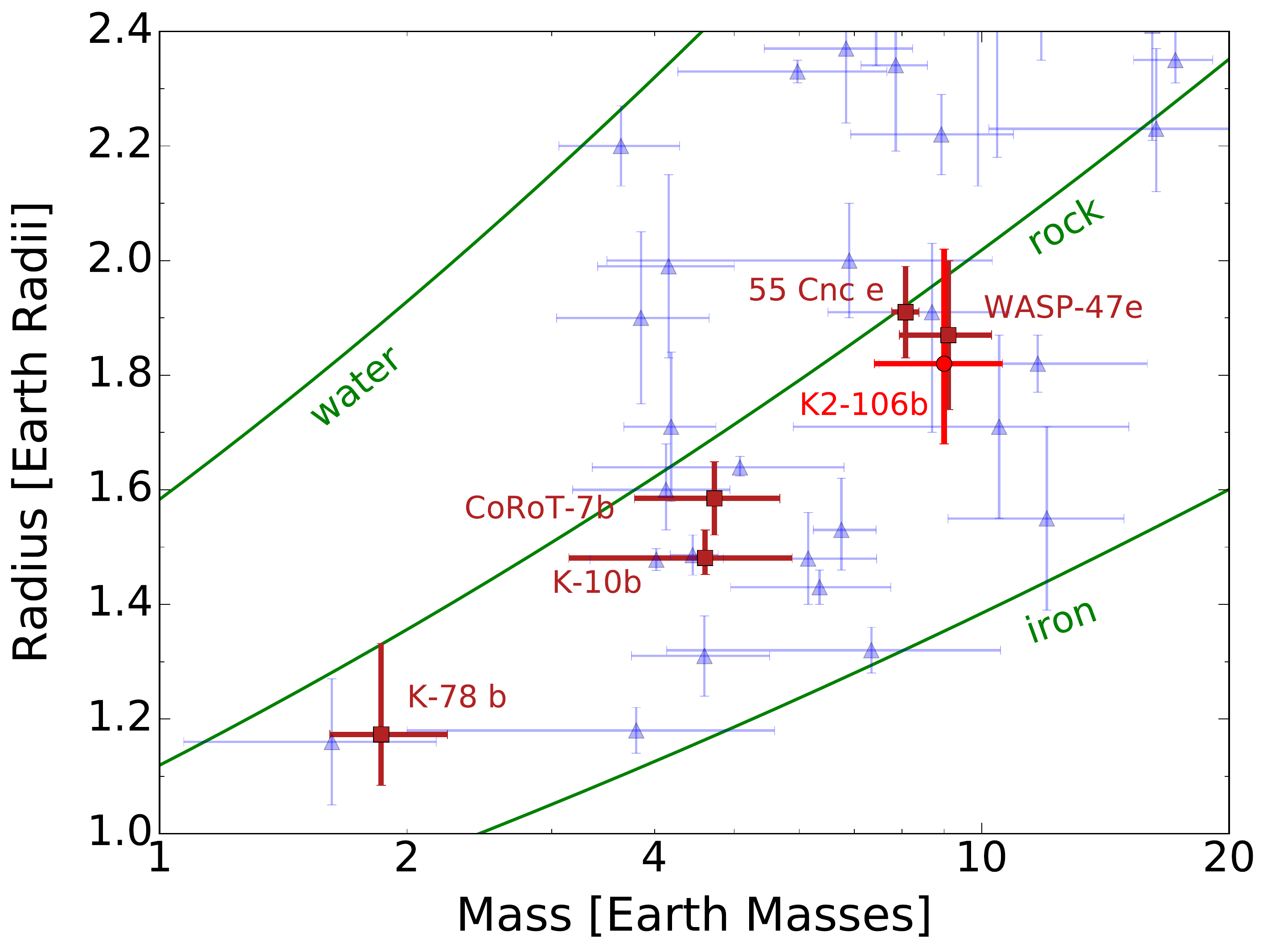}
         }
    \caption{\textit{(a)} Masses and radii of all confirmed planets whose mass and radius are measured to better than 50\% (2$\sigma$) precision (blue triangles). Solar System planets are represented as black squares. Red circles indicate our measurements of \ktwobai{b} and \ktwohbc{b}. Dark red squares represent other USP measurements from the literature. Green curves show the expected planet mass-radius curves for 100\% iron, 100\% rock (Mg$_{\rm2}$SiO$_{\rm4}$), and 100\% water (ice) compositions according to models by \citet{Fortney07}. \textit{(b)} A zoomed in look of the top panel.  The five well-characterized USPs all have masses and radii consistent with mostly rocky compositions and little to no gaseous envelopes.
     }%
   \label{fig:MR}
\end{figure}

The derived planet properties for \ktwobai and \ktwohbc are listed in Tables \ref{tb:params_bai} and \ref{tb:params_hbc} respectively.  Figure \ref{fig:MR_all} shows the masses and radii of \ktwobai{b} and \ktwohbc{b} along with all other planets smaller than 4\,\rearth, whose masses and radii are each known to better than 50\% precision\footnote{NASA Exoplanet Archive, UT 08 February 2017, http://exoplanetarchive.ipac.caltech.edu}. Here we discuss possible planet bulk compositions.

\subsubsection{\ktwobai}	
 
For \ktwobai{b}, we measure a radius \Rp = \baibrp\,\rearth, and a mass \Mp = \baibmp\,\mearth, corresponding to bulk density \rhop = \baibrhop\,\gcc.  It is one of the most massive planets between 2 and 3 \rearth, and likely has a massive heavy-element core.  The compositions of planets in this region of the mass radius diagram are not uniquely determined and could be a range of different admixtures of various chemical species including iron, rock, water and H/He \citep{Rogers10, Valencia13}.  To assess possible compositions, we considered a couple of different two-layer planet models and in each case we constrained the mass fraction of each layer.   

First, we assumed an Earth-composition core (33\% iron, 67\% rock) surrounded by a solar-composition H/He envelope.  We used the work of \citet{Lopez14}, who started with a sample of 1--20\,\mearth cores surrounded by H/He envelopes that are 0.1--50\%  of the total planet mass and recorded the evolution of planet radius and envelope mass over a range of incident fluxes.  Their models consist of planet radii (\Rp) computed over a 4-D grid of planet core mass (\mcore), planet envelope mass (\menv), age, and incident stellar flux (\sinc), i.e. \Rp = \Rp(\mcore, \menv, age, \sinc).  Following \citet{Petigura17}, we interpolated this grid to convert our measured \Mp, \Rp, \sinc, and age into a core mass (envelope mass).  We generated probability distributions for core mass fraction (CMF) by randomly sampling the posteriors of \Mp, \Rp, and \sinc, assuming an age of 5 Gyr.  Varying the age between 3--8 Gyr had negligible effect, which is explained by the fact that at Gyr ages, there is little dependence on age as the heating/cooling budget is close to a steady state value.  From the resulting probability distribution, we constrain CMF $>$ 0.96 and \mcore $>$ 10.8 \mearth at 99.7\% confidence (3$\sigma$).  One potential limitation of our method is that the \citet{Lopez14} models assume the planet incident flux is constant.   However, the luminosity of \ktwobai has increased by a factor of $\sim$2 since evolving off of the main sequence and therefore the planet incident flux was twice as low for most of its lifetime.  Nevertheless, when we repeated this analysis using half the incident flux, the 3$\sigma$ lower limit on the CMF changes by a negligible amount, from 0.96 to 0.95. We conclude that if the planet consists of a H/He envelope atop an Earth-composition core, the envelope is $<$5\% of the planet's mass and the core is $>$10.8 \mearth.  If the iron mass fraction is larger (smaller) than that of Earth, then the planet would need a more (less) extended H/He atmosphere to maintain the same radius.

We also considered a composition of rock (Mg$_{\rm2}$SiO$_{\rm4}$) and water ice.  We randomly drew 100,000 planet masses and radii from the posterior distributions, and converted them into a rock-mass-fraction (RMF) using Equation 7 of \citet{Fortney07}.   From the resulting distribution of RMFs, we conclude that if the planet is indeed a mixture of rock and water ice, then RMF $>$ 81\% at 68.3\% confidence (1$\sigma$).  Moreover, the total mass of rock \mrock $>$ 16\,\mearth at 68.3\% confidence and the planet is denser than pure rock at 39\% confidence.

\subsubsection{\ktwohbc}	

\label{sec:composition_K2106}

For the USP planet \ktwohbc{b}, we measure radius, mass, and density \Rp = \hbcbrp\,\rearth,  \Mp = \hbcbmp\,\mearth, and \rhop = \hbcbrhop\,\gcc.  These are consistent with an Earth-like composition.   Assuming the planet is a mixture of iron and rock, we used Equation 8 of \citet{Fortney07} to convert our mass and radius posteriors into an iron mass fraction (IMF) probability distribution.  The median IMF is 19\% with a 1$\sigma$ upper limit of 33\%, consistent with an Earth-like composition.  With an extremely high incident flux of \hbcbfinc\,\searth, and equilibrium temperature of \hbcbteq\,K, \ktwohbc{b} is the hottest sub-Neptune with a measured density.  At such close proximity to the star, any volatiles would likely have been lost by photoevaporation, leaving a bare $\sim$ 9 \mearth core.

%


The measured radii of planets b and c are larger than those reported by \citet{Adams16} at the $\sim$ 2.5$\sigma$ and $\sim$ 1$\sigma$ level respectively.  \citet{Adams16} measure \Rp = 1.46 $\pm$ 0.14\,\rearth for planet b and \Rp = 2.53 $\pm$ 0.14\,\rearth for planet c.  Adopting their measured radius for planet b with our measured mass yields an iron mass fraction, IMF = 0.8 $\pm$ 0.2.  Although such a large IMF is unlikely based on simulations of planet formation \citep[e.g.][]{Marcus10}, we investigated the source of the measurement discrepancy.  We discovered that \citet{Adams16} underestimate the stellar radius due to an unreported error in the \Teff-\Rstar relations of \citet{Boyajian12}, which they used to convert their spectroscopically measured \Teff (5590 $\pm$ 51 K) into a radius.  Equation 8 of \citet{Boyajian12} was reported as being a third-order polynomial fit to a sample of 33 K--M-dwarfs with precisely measured radii and \Teff.  Equation 9 was reported as a second polynomial fit that extends to hotter temperatures by including the Sun.  However, these equations seem to have been mistakenly swapped --- the polynomial coefficients in Equation 8 belong in Equation 9 and vice-versa.  This can be seen by computing \Rstar(5778\,K) = 1.00 and 0.86 \rsun for Equations 8 and 9 respectively.  The two equations diverge as \Teff exceeds $\sim$ 5300\,K, which is particularly problematic. \citet{Adams16} used Equation 9 to compute \Rstar = 0.83 \rsun but would have computed \Rstar = 0.91 \rsun if they had used Equation 8, which is consistent with our measurement.  Although Equation 8 is preferred for \Teff $\gtrsim$ 5500\,K, neither are particularly reliable for this temperature regime---the Sun is the only fitted data point beyond 5500\,K, which is also where \Rstar and \Teff become significantly age-dependent because of main sequence evolution.  We encourage the authors of any studies who have used Equations 8 and 9 of \citet{Boyajian12} to verify their results.  T. Boyajian has confirmed the error and is working to publish an erratum.

We note that the \Teff and \logg measured by \citet{Adams16} are higher than our measurements.  Our spectroscopic parameters for K2-106 are derived from SME, which has been well-validated by asteroseismically characterized stars \citep{Brewer15}.  Nevertheless, even if we run the \texttt{isochrones} Python package assuming the \Teff, \logg, and \fe values from \citet{Adams16}, we measure stellar parameters \Mstar = 0.96\,\msun and \Rstar = 0.90\,\msun, which are within our measurement errors.  



\subsection{Photoevaporation Desert}

\begin{figure}
   \includegraphics[width=1.0\columnwidth]{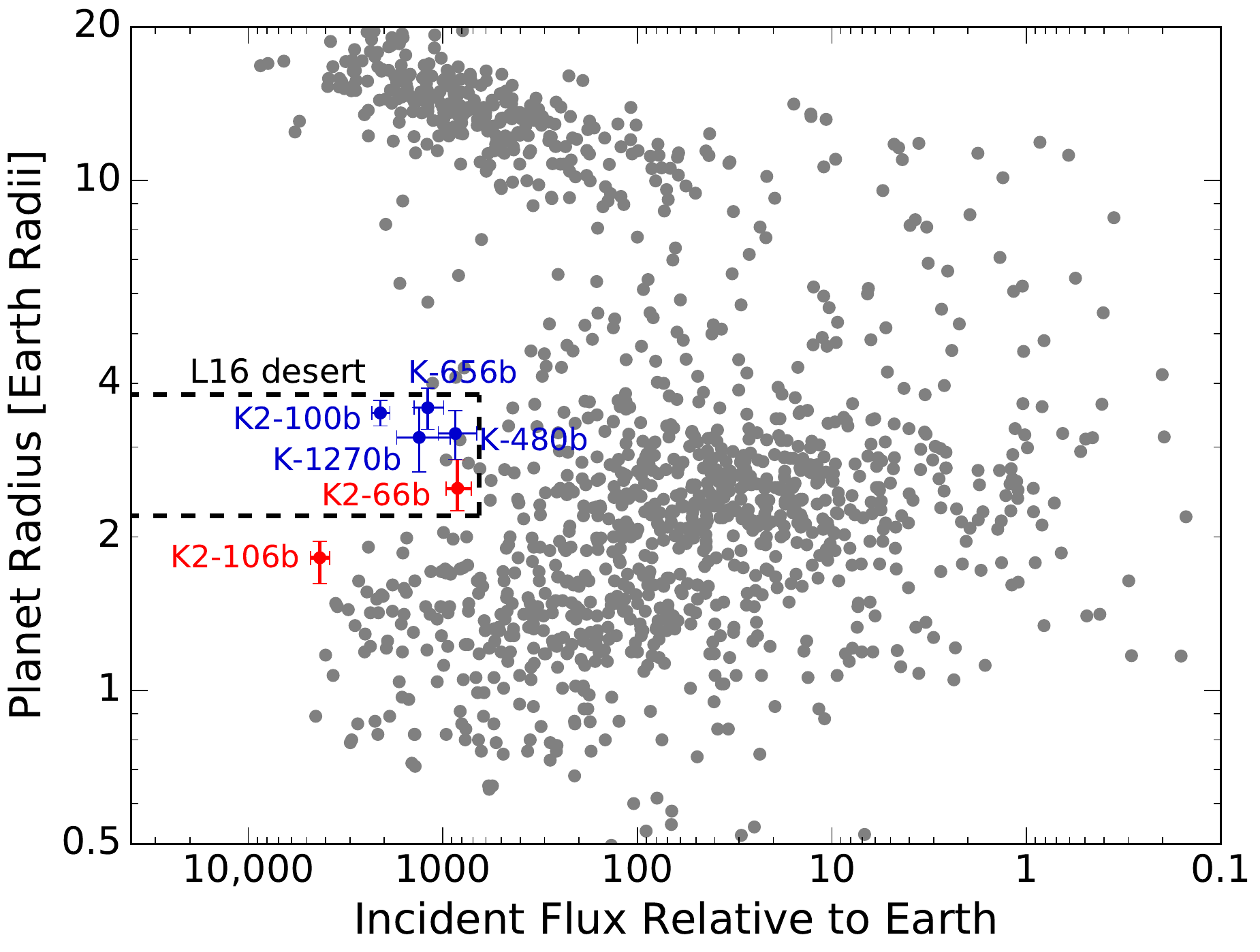}
    \caption{Radii and incident fluxes of all confirmed planets from the NASA Exoplanet Archive. \ktwobai{b} and \ktwohbc{b} are shown in red.  The black dashed box encloses the region of parameter space found by citet{Lundkvist16} to completely lack planets, which we refer to as the L16 desert. \ktwobai{b}, as well as three other planets (blue) occupy the L16 desert and have host stars characterized by both spectroscopic and AO observations.  Four of these five planets have host stars with super-solar luminosities. \ktwohbc{b} is one of the hottest sub-Neptunes found to date.
    } 
   \label{fig:rp_vs_flux}
\end{figure}

The radius and temperature of \ktwobai{b} and \ktwohbc{b} constitute the extremes of planet parameter space.  Figure \ref{fig:rp_vs_flux} shows the radius and incident flux of confirmed planets from the NASA Exoplanet Archive\footnote{NASA Exoplanet Archive, UT 15 February 2017, http://exoplanetarchive.ipac.caltech.edu} (NEA).   \ktwohbc{b} ranks among the hottest sub-Neptunes found to date.  There is a clear absence of very hot planets larger than $\sim$ 2\,\rearth.  Another noticeable feature is that hotter giant planets tend to have larger radii--- the reason for which is highly debated \citep[see][and references therein]{Ginsburg15}.  It would be interesting to see if any trends exist for the larger sub-Neptunes of similar temperature.  \ktwobai{b} occupies the region of parameter space found to be completely devoid of planets by \citet{Lundkvist16} (2.2\,\rearth < \Rp <  3.8\,\rearth, \sinc > 650\,\searth), hereafter referred to as the ``L16 desert''.   

We find that seven other planets fall within the L16 desert.   To assess the reliability of these seven measurements, we examined constraints on the host stellar parameters from spectroscopic and imaging observations.   None of them were asteroseismically characterized by \citet{Lundkvist16}.   According to the Exoplanet Follow-up Observing Program (ExoFOP) database\footnote{https://exofop.ipac.caltech.edu/cfop.php}, five of these stars (K2-100, Kepler-480, Kepler-536, Kepler-656, and Kepler-1270) have properties constrained from spectroscopy and AO imaging.   One of these five stars, Kepler-536, has a stellar companion at $0\farcs56$ separation.  The planet in this system would be much larger than 4\,\rearth if it orbits the companion star rather than the primary \citep{Law14, Furlan17} so we deem this measurement unreliable.   We consider the planet parameters for the other four systems to be reliable and confirm that planets remain in the L16 desert when spectroscopic stellar parameters are adopted.  For K2-100, we adopt the stellar and planet parameters reported in \citep{Mann17}.   The star is a late F dwarf in the 800 Myr Praesepe Cluster.  For Kepler 480, Kepler-656, and Kepler-1270, we had previously obtained HIRES spectra and used the SpecMatch algorithm \citep{Petigura15b} to derive \Teff, \logg, and \fe.  We computed stellar masses and radii using the \texttt{isochrones} package (see \S\ref{sec:stellar_par}).  We find that Kepler-480 is an F8 dwarf, Kepler-1270 is a K1 subgiant, and K2-656 is a high-metallicity G dwarf (\fe = 0.23 $\pm$ 0.05 dex).   The planets in the L16 desert that orbit these four spectroscopically characterized host stars are plotted as blue points and labeled in Figure \ref{fig:rp_vs_flux}.

We examine whether the five planets in the L16 desert share common properties that can be linked to their origins.  First, we note that none of them are USPs --- they have orbital periods of 1.3--6.0 days.   Moreover, four of the five host stars have luminosities $L$ $>$ 1.7\,\lsun.   Based on these two observations, we speculate that planets in the L16 desert are 2--4\,\rearth cores of larger planets that were stripped of their gaseous envelopes by means of photoevaporation.  Such 2--4\,\rearth cores would have higher surface gravities and orbit further from the star than the smaller cores of USPs.  Therefore, the removal of their envelopes by photoevaporation would require stars that are systematically more luminous than USP hosts, consistent with observations.  Mass measurements of other planets in the L16 desert are needed to test the hypothesis that they are cores surrounded by little to no gas.

Given that \ktwobai is a subgiant star, we consider the evolution of the planet's irradiance since the star left the main sequence.  According to Dartmouth stellar evolution models, a star with mass \Mstar  = 1.1\,\msun and \fe = 0.05 dex would have had a radius \Rstar $\approx$ 1.1 \rsun during its main sequence lifetime and have luminosity \Lstar $\approx$ 1.5 \lsun.   Its current luminosity is $\approx$ 3.0 \lsun, meaning that the planet incident flux has increased twofold, from $\approx$ 420 to 840\,\searth since the main sequence era.   This would have boosted the rate of photoevaporation of low-density volatiles in the planet's envelope.  Alternatively, \bai might have formed in a gas-poor disk, preventing it from accumulating much H/He. 

If \ktwobai{b} was stripped of its envelope as the star became a subgiant, then the rapid post-main sequence evolution explains the lack of known planets similar in size and density.  Perhaps we are catching a glimpse of a planet from a population that quickly spirals into their host stars as they evolve off the main sequence \citep[e.g. KELT-8b,][]{Fulton15}.  To test this scenario, we computed an inspiral time, $t_{\mathrm{inspiral}}$ $\approx$ 370 Gyr for K2-66b using Equation 1 of \citet{Lai12} assuming a nominal reduced tidal quality factor $Q^\prime_\star$ = 10$^{7}$.   We conclude that the planet is not on the verge of spiraling into its host star.



\subsection{Ultra-short-period Planets}

Only five other USPs have measured masses and densities:  55 Cnc \citep{Fischer08, Dawson10, Nelson14, Demory16}, CoRoT-7b \citep{Leger09, Bruntt10, Haywood14}, Kepler-10b \citep{Batalha11, Esteves15},  Kepler-78b \citep{Howard13, Pepe13, Grunblatt15}, and WASP-47e \citep{Becker15, Dai15, Sinukoff17a}.  These planets are plotted on the mass-radius diagram in Figure \ref{fig:MR_USP}.  The properties of these planets and their host stars are provided in Table \ref{tb:USP}. All of them have masses and radii consistent with admixtures of rock and iron with little to no surrounding volatiles.  This is consistent with the notion that USPs are the remnant cores of larger planets that lost their gaseous envelopes or formed without much gas in the first place.  It is curious that three of the six well-characterized USPs have consistent masses and radii that are $\sim$ 1.7--2.0\,\rearth and $\sim$ 8--10\,\mearth.   Perhaps these planets constitute an upper size and mass limit to the cores of the larger planets from which they form.  If all USPs have similar rocky compositions, then the observed absence of USPs $>$2.0\,\rearth naturally translates to an upper mass limit.  Some sub-Neptune-size planets with $P$ $>$ 1 day have cores $>$ 10\,\mearth (e.g. \ktwobai{b}), but there are no such examples of USPs.  More well-characterized USPs are needed to reveal their core mass distribution.  

We note that the three well-characterized USPs with $\sim$ 8--10\,\mearth cores (\ktwohbc{b}, 55 Cnc e, WASP-47e) have host stars with super-solar metallicities, whereas two of the three well-characterized USPs with masses $\lesssim$ 5\,\mearth (Kepler-78b and Kepler-10b) have host stars with sub-solar metallicities.   With only six data points, a correlation cannot be claimed, but this motivates a more complete analysis of all USPs beyond the scope of this study.

USPs are unlikely to be remnants of hot-Jupiters.  While earlier studies argued that USPs could be the leftover cores of hot-Jupiters that experienced Roche lobe overflow \citep[RLO, e.g.][]{Valsecchi14}, simulations by \citet{Valsecchi15} and \citet{Jackson16} suggest that RLO of planets with cores $\lesssim$ 10\,\mearth would tend to expand their orbits to $P$ $>$ 1 day.   Moreover, \citet{Winn17} found that the \fe distribution of USP host stars is inconsistent with that of hot-Jupiter host stars, and consistent with that of stars hosting hot planets of Neptune-size or smaller.  This suggests the that the majority of USPs are not remnants of hot-Jupiters but could be remnants of Neptune- or sub-Neptune-size planets.

Five of the six well-characterized USPs have known planetary companions.  The single exception is Kepler-78b, which orbits an active star, hampering the ability to detect planets with longer orbital periods.   The number of detected companions to USPs is consistent with a 50--100\% occurrence rate of additional planets $P$ < 45 days, depending on the assumed distribution of mutual inclinations and assuming 100\% detection completeness \citep{Adams16}.   

It remains unclear how USPs settle so close to their host stars, but the multiplicity of these systems ($P$ $<$ 50 days) hints that they form via inward migration mechanisms involving multiple planets.   For example, \citet{Hansen15} demonstrated that tidal decay of 55 Cnc e from beyond its current orbit would have sent the planet through multiple secular resonances, exciting its orbital eccentricity and inclination.  A shrinking periastron distance would subsequently boost tidal evolution and increase the rate of orbital decay.  However, unless the perturber has a mass comparable to Jupiter, secular interactions are usually too weak to overcome relativistic precession at short orbital periods \citep{Lee17}.  Thus, secular interactions can only explain USP systems that also host close-in giant planets like 55 Cnc and WASP-47.  Alternatively, USPs might have migrated through a gas disk to their current orbits via mean motion resonances (MMRs) with other planets.   However, companions of USPs detected to date are not in MMR.  It is possible that resonant companions were engulfed by the star or collided to form a single object.  Formation of USPs via MMR would require the disk to extend very close to the star. USPs could also have been gravitationally scattered inwards by another companion, but this is difficult to reconcile with the observed presence of multiple companions on close-in orbits, which would be unstable at modest eccentricities.  \citet{Lee17} show that the observed USP population is consistent with in-situ formation or disk migration followed by tidal migration. Any complete theory of planet formation must account for the presence of these rocky $\sim$ 5-10 \mearth USPs with close neighbors. 

\begin{deluxetable*}{lrrrrrrrrr}[h!]
\tablecaption{Ultra-short-period planets with measured masses.}
\tablehead{\colhead{Name}  & \colhead{\Mstar} & \colhead{\Rstar} & \colhead{\fe} & \colhead{P} & \colhead{\Rp} & \colhead{\Mp} & \colhead{$\rhop$}  & \colhead{$N_{\rm{pl}}$} & \colhead{References} \\
& \colhead{(\msun)} & \colhead{(\rsun)} & \colhead{(dex)} & \colhead{(days)} & \colhead{(\rearth)} & \colhead{(\mearth)} & \colhead{(\gcc)}  &  & }
\startdata
55 Cnc e & 0.905 $\pm$ 0.015 & 0.943 $\pm$ 0.010 & 0.31 $\pm$ 0.04 & 0.74 & 1.92 $\pm$ 0.08 & 8.08 $\pm$ 0.31 & 6.3$^{+0.8}_{-0.7}$ & 5 &  \valenti, \vonbraun, \demory  \\
CoRot-7b & 0.91 $\pm$ 0.03 & 0.82 $\pm$ 0.04 & 0.12 $\pm$ 0.06 & 0.85 & 1.585 $\pm$ 0.064 & 4.73 $\pm$ 0.95 & 6.61 $\pm$ 1.33 & 2 &  \leger, \bruntt, \haywood \\
Kepler-10b & 0.913 $\pm$ 0.022 & 1.065 $\pm$ 0.009 & $-0.15$ $\pm$ 0.04 & 0.84 & 1.48$^{+0.05}_{-0.03}$& 4.61$^{+1.27}_{-1.46}$ & 8.0 $\pm$ 3.0 & 2 & \batalha, \esteves \\
Kepler-78b & 0.83 $\pm$ 0.05 & 0.74 $\pm$ 0.05 & -0.08 $\pm$ 0.04 & 0.36 & 1.18$^{+0.16}_{-0.09}$ & 1.86$^{+0.38}_{-0.25}$ & 5.57$^{+3.02}_{-1.31}$ & 1 &  \ojeda, \howard, \pepe \\
WASP-47e & 0.99 $\pm$ 0.05 & 1.18 $\pm$ 0.08 & 0.36 $\pm$ 0.05 & 0.79 & 1.87 $\pm$ 0.13 & 9.11 $\pm$ 1.17 & 7.63 $\pm$ 1.90 & 4 & \becker, \sinukoff \\
\ktwohbc{b} & \hbcstarrad &  \hbcstarmass &  \hbcstarfe & 0.57 & \hbcbrp & \hbcbmp & \hbcbrhop & 2 & \thisstudy 
\enddata
\tablecomments{\valenti: \citet{Valenti05}, \vonbraun: \citet{VonBraun11}, \demory: \citet{Demory16}, \leger: \citet{Leger09}, \bruntt: \citet{Bruntt10}, \haywood: \citet{Haywood14}, \batalha: \citet{Batalha11}, \esteves: \citet{Esteves15}, \ojeda: \citet{Ojeda13}, \howard:  \citet{Howard13}, \pepe: \citet{Pepe13},  \becker: \citet{Becker15}, \sinukoff: \citet{Sinukoff17a}.}
\label{tb:USP}
\end{deluxetable*}

\section{Conclusion}
\label{sec:conclusion}
We have measured the masses and densities of two extremely hot sub-Neptunes, \ktwobai{b} and \ktwohbc{b}.  We have characterized their stellar hosts using high-resolution spectroscopy and adaptive optics imaging. The radius of \ktwobai{b}, \Rp = \baibrp\,\rearth measured from \ktwo photometry and mass, \Mp = \baibmp\,\mearth measured from Keck-HIRES RVs are consistent with a mostly rocky composition and little to no low-density volatiles, making it one of the densest planets of its size.  It is one of the few known planets in the ``photoevaporation desert" (\Rp = 2.2--3.8\,\rearth, \sinc $\ge$ 650\,\searth), and the first such planet with a measured mass.  These planets tend to orbit stars more luminous than the Sun, which suggests that they might have systematically higher densities due to increased photoevaporation. The measured radius, \Rp = \hbcbrp\,\rearth and mass, \Mp = \baibmp\,\mearth of \ktwohbc{b} indicate an Earth-like composition, similar to the four other USPs with measured densities.  It is the hottest sub-Neptune with a measured mass, and could be the stripped core of a more massive planet.  \ktwobai{b} and \ktwohbc{b} join the rare class of planets larger than 1.5 \rearth with mostly rocky compositions.

\acknowledgements
We thank the many observers who contributed to the measurements reported here. We gratefully acknowledge the efforts and dedication of the Keck Observatory staff.  We thank Tabetha Boyajian for helpful discussions. This paper includes data collected by the \ktwo mission. Funding for the \ktwo mission is provided by the NASA Science Mission directorate.  E.~S.\ is supported by a post-graduate scholarship from the Natural Sciences and Engineering Research Council of Canada. E.~A.~P.\ acknowledges support by NASA through a Hubble Fellowship grant awarded by the Space Telescope Science Institute, which is operated by the Association of Universities for Research in Astronomy, Inc., for NASA, under contract NAS 5-26555. B.~J.~F.\ was supported by the National Science Foundation Graduate Research Fellowship under grant No. 2014184874.  A.~W.~H.\ acknowledges support for our \ktwo team through a NASA Astrophysics Data Analysis Program grant.  A.~W.~H.\ and I.~J.~M.~C.\ acknowledge support from the \ktwo Guest Observer Program.  L.~M.~W.\ acknowledges the Trottier Family Foundation for their generous support. J.~R.~C.\ acknowledges support from the Kepler Participating Scientist program (NNX14AB85G). This work was performed [in part] under contract with the Jet Propulsion Laboratory (JPL) funded by NASA through the Sagan Fellowship Program executed by the NASA Exoplanet Science Institute. This research has made use of the NASA Exoplanet Archive, which is operated by the California Institute of Technology, under contract with the National Aeronautics and Space Administration under the Exoplanet Exploration Program. Finally, the authors extend special thanks to those of Hawai`ian ancestry on whose sacred mountain of Maunakea we are privileged to be guests.  Without their generous hospitality, the Keck observations presented herein would not have been possible.

{\it Facilities:} \facility{Kepler}, \facility{Keck-HIRES}. 
\FloatBarrier
\bibliographystyle{apj}
\bibliography{references}

\begin{thebibliography}{}
\expandafter\ifx\csname natexlab\endcsname\relax\def\natexlab#1{#1}\fi

\bibitem[{{Adams} {et~al.}(2016{\natexlab{a}}){Adams}, {Jackson}, \&
  {Endl}}]{Adams16pig}
{Adams}, E.~R., {Jackson}, B., \& {Endl}, M. 2016{\natexlab{a}}, \aj, 152, 47

\bibitem[{{Adams} {et~al.}(2016{\natexlab{b}}){Adams}, {Jackson}, {Endl},
  {Cochran}, {MacQueen}, {Duev}, {Jensen-Clem}, {Salama}, {Ziegler}, {Baranec},
  {Kulkarni}, {Law}, \& {Riddle}}]{Adams16}
{Adams}, E.~R., {Jackson}, B., {Endl}, M., {et~al.} 2016{\natexlab{b}}, ArXiv
  e-prints, arXiv:1611.00397

\bibitem[{{Agol} {et~al.}(2005){Agol}, {Steffen}, {Sari}, \&
  {Clarkson}}]{Agol05}
{Agol}, E., {Steffen}, J., {Sari}, R., \& {Clarkson}, W. 2005, \mnras, 359, 567

\bibitem[{{Barros} {et~al.}(2016){Barros}, {Demangeon}, \&
  {Deleuil}}]{Barros16}
{Barros}, S.~C.~C., {Demangeon}, O., \& {Deleuil}, M. 2016, \aap, 594, A100

\bibitem[{{Batalha} {et~al.}(2011){Batalha}, {Borucki}, {Bryson}, {Buchhave},
  {Caldwell}, {Christensen-Dalsgaard}, {Ciardi}, {Dunham}, {Fressin},
  {Gautier}, {Gilliland}, {Haas}, {Howell}, {Jenkins}, {Kjeldsen}, {Koch},
  {Latham}, {Lissauer}, {Marcy}, {Rowe}, {Sasselov}, {Seager}, {Steffen},
  {Torres}, {Basri}, {Brown}, {Charbonneau}, {Christiansen}, {Clarke},
  {Cochran}, {Dupree}, {Fabrycky}, {Fischer}, {Ford}, {Fortney}, {Girouard},
  {Holman}, {Johnson}, {Isaacson}, {Klaus}, {Machalek}, {Moorehead},
  {Morehead}, {Ragozzine}, {Tenenbaum}, {Twicken}, {Quinn}, {VanCleve},
  {Walkowicz}, {Welsh}, {Devore}, \& {Gould}}]{Batalha11}
{Batalha}, N.~M., {Borucki}, W.~J., {Bryson}, S.~T., {et~al.} 2011, \apj, 729,
  27

\bibitem[{{Becker} {et~al.}(2015){Becker}, {Vanderburg}, {Adams}, {Rappaport},
  \& {Schwengeler}}]{Becker15}
{Becker}, J.~C., {Vanderburg}, A., {Adams}, F.~C., {Rappaport}, S.~A., \&
  {Schwengeler}, H.~M. 2015, \apjl, 812, L18

\bibitem[{{Boyajian} {et~al.}(2012){Boyajian}, {von Braun}, {van Belle},
  {McAlister}, {ten Brummelaar}, {Kane}, {Muirhead}, {Jones}, {White},
  {Schaefer}, {Ciardi}, {Henry}, {L{\'o}pez-Morales}, {Ridgway}, {Gies}, {Jao},
  {Rojas-Ayala}, {Parks}, {Sturmann}, {Sturmann}, {Turner}, {Farrington},
  {Goldfinger}, \& {Berger}}]{Boyajian12}
{Boyajian}, T.~S., {von Braun}, K., {van Belle}, G., {et~al.} 2012, \apj, 757,
  112

\bibitem[{{Brewer} {et~al.}(2015){Brewer}, {Fischer}, {Basu}, {Valenti}, \&
  {Piskunov}}]{Brewer15}
{Brewer}, J.~M., {Fischer}, D.~A., {Basu}, S., {Valenti}, J.~A., \& {Piskunov},
  N. 2015, \apj, 805, 126

\bibitem[{{Brewer} {et~al.}(2016){Brewer}, {Fischer}, {Valenti}, \&
  {Piskunov}}]{Brewer16}
{Brewer}, J.~M., {Fischer}, D.~A., {Valenti}, J.~A., \& {Piskunov}, N. 2016,
  \apjs, 225, 32

\bibitem[{{Bruntt} {et~al.}(2010){Bruntt}, {Deleuil}, {Fridlund}, {Alonso},
  {Bouchy}, {Hatzes}, {Mayor}, {Moutou}, \& {Queloz}}]{Bruntt10}
{Bruntt}, H., {Deleuil}, M., {Fridlund}, M., {et~al.} 2010, \aap, 519, A51

\bibitem[{{Burke} {et~al.}(2015){Burke}, {Christiansen}, {Mullally}, {Seader},
  {Huber}, {Rowe}, {Coughlin}, {Thompson}, {Catanzarite}, {Clarke}, {Morton},
  {Caldwell}, {Bryson}, {Haas}, {Batalha}, {Jenkins}, {Tenenbaum}, {Twicken},
  {Li}, {Quintana}, {Barclay}, {Henze}, {Borucki}, {Howell}, \&
  {Still}}]{Burke15}
{Burke}, C.~J., {Christiansen}, J.~L., {Mullally}, F., {et~al.} 2015, \apj,
  809, 8

\bibitem[{{Butler} {et~al.}(1996){Butler}, {Marcy}, {Williams}, {McCarthy},
  {Dosanjh}, \& {Vogt}}]{Butler96}
{Butler}, R.~P., {Marcy}, G.~W., {Williams}, E., {et~al.} 1996, \pasp, 108, 500

\bibitem[{{Carter} {et~al.}(2012){Carter}, {Agol}, {Chaplin}, {Basu},
  {Bedding}, {Buchhave}, {Christensen-Dalsgaard}, {Deck}, {Elsworth},
  {Fabrycky}, {Ford}, {Fortney}, {Hale}, {Handberg}, {Hekker}, {Holman},
  {Huber}, {Karoff}, {Kawaler}, {Kjeldsen}, {Lissauer}, {Lopez}, {Lund},
  {Lundkvist}, {Metcalfe}, {Miglio}, {Rogers}, {Stello}, {Borucki}, {Bryson},
  {Christiansen}, {Cochran}, {Geary}, {Gilliland}, {Haas}, {Hall}, {Howard},
  {Jenkins}, {Klaus}, {Koch}, {Latham}, {MacQueen}, {Sasselov}, {Steffen},
  {Twicken}, \& {Winn}}]{Carter12}
{Carter}, J.~A., {Agol}, E., {Chaplin}, W.~J., {et~al.} 2012, Science, 337, 556

\bibitem[{{Ciardi} {et~al.}(2015){Ciardi}, {Beichman}, {Horch}, \&
  {Howell}}]{Ciardi15}
{Ciardi}, D.~R., {Beichman}, C.~A., {Horch}, E.~P., \& {Howell}, S.~B. 2015,
  \apj, 805, 16

\bibitem[{{Crepp} {et~al.}(2012){Crepp}, {Johnson}, {Howard}, {Marcy},
  {Fischer}, {Hillenbrand}, {Yantek}, {Delaney}, {Wright}, {Isaacson}, \&
  {Montet}}]{Crepp12}
{Crepp}, J.~R., {Johnson}, J.~A., {Howard}, A.~W., {et~al.} 2012, \apj, 761, 39

\bibitem[{{Crossfield} {et~al.}(2016){Crossfield}, {Ciardi}, {Petigura},
  {Sinukoff}, {Schlieder}, {Howard}, {Beichman}, {Isaacson}, {Dressing},
  {Christiansen}, {Fulton}, {L{\'e}pine}, {Weiss}, {Hirsch}, {Livingston},
  {Baranec}, {Law}, {Riddle}, {Ziegler}, {Howell}, {Horch}, {Everett}, {Teske},
  {Martinez}, {Obermeier}, {Benneke}, {Scott}, {Deacon}, {Aller}, {Hansen},
  {Mancini}, {Ciceri}, {Brahm}, {Jord{\'a}n}, {Knutson}, {Henning}, {Bonnefoy},
  {Liu}, {Crepp}, {Lothringer}, {Hinz}, {Bailey}, {Skemer}, \&
  {Defrere}}]{Crossfield16}
{Crossfield}, I.~J.~M., {Ciardi}, D.~R., {Petigura}, E.~A., {et~al.} 2016,
  \apjs, 226, 7

\bibitem[{{Dai} {et~al.}(2015){Dai}, {Winn}, {Arriagada}, {Butler}, {Crane},
  {Johnson}, {Shectman}, {Teske}, {Thompson}, {Vanderburg}, \&
  {Wittenmyer}}]{Dai15}
{Dai}, F., {Winn}, J.~N., {Arriagada}, P., {et~al.} 2015, \apjl, 813, L9

\bibitem[{{Dawson} \& {Fabrycky}(2010)}]{Dawson10}
{Dawson}, R.~I., \& {Fabrycky}, D.~C. 2010, \apj, 722, 937

\bibitem[{{Demory} {et~al.}(2016){Demory}, {Gillon}, {Madhusudhan}, \&
  {Queloz}}]{Demory16}
{Demory}, B.-O., {Gillon}, M., {Madhusudhan}, N., \& {Queloz}, D. 2016, \mnras,
  455, 2018

\bibitem[{{Dotter} {et~al.}(2008){Dotter}, {Chaboyer}, {Jevremovi{\'c}},
  {Kostov}, {Baron}, \& {Ferguson}}]{Dotter08}
{Dotter}, A., {Chaboyer}, B., {Jevremovi{\'c}}, D., {et~al.} 2008, \apjs, 178,
  89

\bibitem[{{Dressing} {et~al.}(2015){Dressing}, {Charbonneau}, {Dumusque},
  {Gettel}, {Pepe}, {Collier Cameron}, {Latham}, {Molinari}, {Udry}, {Affer},
  {Bonomo}, {Buchhave}, {Cosentino}, {Figueira}, {Fiorenzano}, {Harutyunyan},
  {Haywood}, {Johnson}, {Lopez-Morales}, {Lovis}, {Malavolta}, {Mayor},
  {Micela}, {Motalebi}, {Nascimbeni}, {Phillips}, {Piotto}, {Pollacco},
  {Queloz}, {Rice}, {Sasselov}, {S{\'e}gransan}, {Sozzetti}, {Szentgyorgyi}, \&
  {Watson}}]{Dressing15a}
{Dressing}, C.~D., {Charbonneau}, D., {Dumusque}, X., {et~al.} 2015, \apj, 800,
  135

\bibitem[{{Eastman} {et~al.}(2013){Eastman}, {Gaudi}, \& {Agol}}]{Eastman13}
{Eastman}, J., {Gaudi}, B.~S., \& {Agol}, E. 2013, \pasp, 125, 83

\bibitem[{{Esteves} {et~al.}(2015){Esteves}, {De Mooij}, \&
  {Jayawardhana}}]{Esteves15}
{Esteves}, L.~J., {De Mooij}, E.~J.~W., \& {Jayawardhana}, R. 2015, \apj, 804,
  150

\bibitem[{{Fischer} {et~al.}(2008){Fischer}, {Marcy}, {Butler}, {Vogt},
  {Laughlin}, {Henry}, {Abouav}, {Peek}, {Wright}, {Johnson}, {McCarthy}, \&
  {Isaacson}}]{Fischer08}
{Fischer}, D.~A., {Marcy}, G.~W., {Butler}, R.~P., {et~al.} 2008, \apj, 675,
  790

\bibitem[{{Ford}(2006)}]{Ford06}
{Ford}, E.~B. 2006, \apj, 642, 505

\bibitem[{{Foreman-Mackey} {et~al.}(2013){Foreman-Mackey}, {Hogg}, {Lang}, \&
  {Goodman}}]{Foreman-Mackey13}
{Foreman-Mackey}, D., {Hogg}, D.~W., {Lang}, D., \& {Goodman}, J. 2013, \pasp,
  125, 306

\bibitem[{{Fortney} {et~al.}(2007){Fortney}, {Marley}, \& {Barnes}}]{Fortney07}
{Fortney}, J.~J., {Marley}, M.~S., \& {Barnes}, J.~W. 2007, \apj, 659, 1661

\bibitem[{{Fressin} {et~al.}(2013){Fressin}, {Torres}, {Charbonneau}, {Bryson},
  {Christiansen}, {Dressing}, {Jenkins}, {Walkowicz}, \& {Batalha}}]{Fressin13}
{Fressin}, F., {Torres}, G., {Charbonneau}, D., {et~al.} 2013, \apj, 766, 81

\bibitem[{{Fulton} {et~al.}(2015){Fulton}, {Collins}, {Gaudi}, {Stassun},
  {Pepper}, {Beatty}, {Siverd}, {Penev}, {Howard}, {Baranec}, {Corfini},
  {Eastman}, {Gregorio}, {Law}, {Lund}, {Oberst}, {Penny}, {Riddle},
  {Rodriguez}, {Stevens}, {Zambelli}, {Ziegler}, {Bieryla}, {D'Ago}, {DePoy},
  {Jensen}, {Kielkopf}, {Latham}, {Manner}, {Marshall}, {McLeod}, \&
  {Reed}}]{Fulton15}
{Fulton}, B.~J., {Collins}, K.~A., {Gaudi}, B.~S., {et~al.} 2015, \apj, 810, 30

\bibitem[{{Furlan} {et~al.}(2017){Furlan}, {Ciardi}, {Everett}, {Saylors},
  {Teske}, {Horch}, {Howell}, {van Belle}, {Hirsch}, {Gautier}, {Adams},
  {Barrado}, {Cartier}, {Dressing}, {Dupree}, {Gilliland}, {Lillo-Box},
  {Lucas}, \& {Wang}}]{Furlan17}
{Furlan}, E., {Ciardi}, D.~R., {Everett}, M.~E., {et~al.} 2017, \aj, 153, 71

\bibitem[{Gelman \& Rubin(1992)}]{Gelman92}
Gelman, A., \& Rubin, D.~B. 1992, Statist. Sci., 7, 457

\bibitem[{{Ginzburg} \& {Sari}(2015)}]{Ginsburg15}
{Ginzburg}, S., \& {Sari}, R. 2015, \apj, 803, 111

\bibitem[{{Goldreich} \& {Soter}(1966)}]{Goldreich66}
{Goldreich}, P., \& {Soter}, S. 1966, \icarus, 5, 375

\bibitem[{Goodman \& Weare(2010)}]{Goodman10}
Goodman, J., \& Weare, J. 2010, Communications in Applied Mathematics and
  Computational Science, 5, 65

\bibitem[{{Grunblatt} {et~al.}(2015){Grunblatt}, {Howard}, \&
  {Haywood}}]{Grunblatt15}
{Grunblatt}, S.~K., {Howard}, A.~W., \& {Haywood}, R.~D. 2015,
  arXiv:1501.00369, arXiv:1501.00369

\bibitem[{{Hadden} \& {Lithwick}(2014)}]{Hadden14}
{Hadden}, S., \& {Lithwick}, Y. 2014, \apj, 787, 80

\bibitem[{{Hansen} \& {Zink}(2015)}]{Hansen15}
{Hansen}, B.~M.~S., \& {Zink}, J. 2015, \mnras, 450, 4505

\bibitem[{{Haywood} {et~al.}(2014){Haywood}, {Collier Cameron}, {Queloz},
  {Barros}, {Deleuil}, {Fares}, {Gillon}, {Lanza}, {Lovis}, {Moutou}, {Pepe},
  {Pollacco}, {Santerne}, {S{\'e}gransan}, \& {Unruh}}]{Haywood14}
{Haywood}, R.~D., {Collier Cameron}, A., {Queloz}, D., {et~al.} 2014, \mnras,
  443, 2517

\bibitem[{{Henning} {et~al.}(2009){Henning}, {O'Connell}, \&
  {Sasselov}}]{Henning09}
{Henning}, W.~G., {O'Connell}, R.~J., \& {Sasselov}, D.~D. 2009, \apj, 707,
  1000

\bibitem[{{Holman} \& {Murray}(2005)}]{Holman05}
{Holman}, M.~J., \& {Murray}, N.~W. 2005, Science, 307, 1288

\bibitem[{{Howard} \& {Fulton}(2016)}]{Howard16}
{Howard}, A.~W., \& {Fulton}, B.~J. 2016, \pasp, 128, 114401

\bibitem[{{Howard} {et~al.}(2009){Howard}, {Johnson}, {Marcy}, {Fischer},
  {Wright}, {Henry}, {Giguere}, {Isaacson}, {Valenti}, {Anderson}, \&
  {Piskunov}}]{Howard09}
{Howard}, A.~W., {Johnson}, J.~A., {Marcy}, G.~W., {et~al.} 2009, \apj, 696, 75

\bibitem[{{Howard} {et~al.}(2010){Howard}, {Johnson}, {Marcy}, {Fischer},
  {Wright}, {Bernat}, {Henry}, {Peek}, {Isaacson}, {Apps}, {Endl}, {Cochran},
  {Valenti}, {Anderson}, \& {Piskunov}}]{Howard10}
---. 2010, \apj, 721, 1467

\bibitem[{{Howard} {et~al.}(2012){Howard}, {Marcy}, {Bryson}, {Jenkins},
  {Rowe}, {Batalha}, {Borucki}, {Koch}, {Dunham}, {Gautier}, {Van Cleve},
  {Cochran}, {Latham}, {Lissauer}, {Torres}, {Brown}, {Gilliland}, {Buchhave},
  {Caldwell}, {Christensen-Dalsgaard}, {Ciardi}, {Fressin}, {Haas}, {Howell},
  {Kjeldsen}, {Seager}, {Rogers}, {Sasselov}, {Steffen}, {Basri},
  {Charbonneau}, {Christiansen}, {Clarke}, {Dupree}, {Fabrycky}, {Fischer},
  {Ford}, {Fortney}, {Tarter}, {Girouard}, {Holman}, {Johnson}, {Klaus},
  {Machalek}, {Moorhead}, {Morehead}, {Ragozzine}, {Tenenbaum}, {Twicken},
  {Quinn}, {Isaacson}, {Shporer}, {Lucas}, {Walkowicz}, {Welsh}, {Boss},
  {Devore}, {Gould}, {Smith}, {Morris}, {Prsa}, {Morton}, {Still}, {Thompson},
  {Mullally}, {Endl}, \& {MacQueen}}]{Howard12}
{Howard}, A.~W., {Marcy}, G.~W., {Bryson}, S.~T., {et~al.} 2012, \apjs, 201, 15

\bibitem[{{Howard} {et~al.}(2013){Howard}, {Sanchis-Ojeda}, {Marcy}, {Johnson},
  {Winn}, {Isaacson}, {Fischer}, {Fulton}, {Sinukoff}, \& {Fortney}}]{Howard13}
{Howard}, A.~W., {Sanchis-Ojeda}, R., {Marcy}, G.~W., {et~al.} 2013, \nat, 503,
  381

\bibitem[{{Howard} {et~al.}(2014){Howard}, {Marcy}, {Fischer}, {Isaacson},
  {Muirhead}, {Henry}, {Boyajian}, {von Braun}, {Becker}, {Wright}, \&
  {Johnson}}]{Howard14}
{Howard}, A.~W., {Marcy}, G.~W., {Fischer}, D.~A., {et~al.} 2014, \apj, 794, 51

\bibitem[{{Howell} {et~al.}(2014){Howell}, {Sobeck}, {Haas}, {Still},
  {Barclay}, {Mullally}, {Troeltzsch}, {Aigrain}, {Bryson}, {Caldwell},
  {Chaplin}, {Cochran}, {Huber}, {Marcy}, {Miglio}, {Najita}, {Smith},
  {Twicken}, \& {Fortney}}]{Howell14}
{Howell}, S.~B., {Sobeck}, C., {Haas}, M., {et~al.} 2014, \pasp, 126, 398

\bibitem[{{Isaacson} \& {Fischer}(2010)}]{Isaacson10}
{Isaacson}, H., \& {Fischer}, D. 2010, \apj, 725, 875

\bibitem[{{Jackson} {et~al.}(2016){Jackson}, {Jensen}, {Peacock}, {Arras}, \&
  {Penev}}]{Jackson16}
{Jackson}, B., {Jensen}, E., {Peacock}, S., {Arras}, P., \& {Penev}, K. 2016,
  Celestial Mechanics and Dynamical Astronomy, 126, 227

\bibitem[{Kass \& Raftery(1995)}]{Kass95}
Kass, R.~E., \& Raftery, A.~E. 1995, Journal of the American Statistical
  Association, 90, 773

\bibitem[{{Kolbl} {et~al.}(2015){Kolbl}, {Marcy}, {Isaacson}, \&
  {Howard}}]{Kolbl15}
{Kolbl}, R., {Marcy}, G.~W., {Isaacson}, H., \& {Howard}, A.~W. 2015, \aj, 149,
  18

\bibitem[{{Kraus} \& {Hillenbrand}(2007)}]{Kraus07}
{Kraus}, A.~L., \& {Hillenbrand}, L.~A. 2007, \aj, 134, 2340

\bibitem[{{Kreidberg}(2015)}]{Kreidberg15}
{Kreidberg}, L. 2015, \pasp, 127, 1161

\bibitem[{{Lai}(2012)}]{Lai12}
{Lai}, D. 2012, \mnras, 423, 486

\bibitem[{{Lainey}(2016)}]{Lainey16}
{Lainey}, V. 2016, Celestial Mechanics and Dynamical Astronomy, 126, 145

\bibitem[{{Law} {et~al.}(2014){Law}, {Morton}, {Baranec}, {Riddle},
  {Ravichandran}, {Ziegler}, {Johnson}, {Tendulkar}, {Bui}, {Burse}, {Das},
  {Dekany}, {Kulkarni}, {Punnadi}, \& {Ramaprakash}}]{Law14}
{Law}, N.~M., {Morton}, T., {Baranec}, C., {et~al.} 2014, \apj, 791, 35

\bibitem[{{Lee} \& {Chiang}(2016)}]{Lee16}
{Lee}, E.~J., \& {Chiang}, E. 2016, \apj, 817, 90

\bibitem[{{Lee} \& {Chiang}(2017)}]{Lee17}
---. 2017, ArXiv e-prints, arXiv:1702.08461

\bibitem[{{L{\'e}ger} {et~al.}(2009){L{\'e}ger}, {Rouan}, {Schneider}, {Barge},
  {Fridlund}, {Samuel}, {Ollivier}, {Guenther}, {Deleuil}, {Deeg}, {Auvergne},
  {Alonso}, {Aigrain}, {Alapini}, {Almenara}, {Baglin}, {Barbieri}, {Bruntt},
  {Bord{\'e}}, {Bouchy}, {Cabrera}, {Catala}, {Carone}, {Carpano}, {Csizmadia},
  {Dvorak}, {Erikson}, {Ferraz-Mello}, {Foing}, {Fressin}, {Gandolfi},
  {Gillon}, {Gondoin}, {Grasset}, {Guillot}, {Hatzes}, {H{\'e}brard}, {Jorda},
  {Lammer}, {Llebaria}, {Loeillet}, {Mayor}, {Mazeh}, {Moutou}, {P{\"a}tzold},
  {Pont}, {Queloz}, {Rauer}, {Renner}, {Samadi}, {Shporer}, {Sotin}, {Tingley},
  {Wuchterl}, {Adda}, {Agogu}, {Appourchaux}, {Ballans}, {Baron}, {Beaufort},
  {Bellenger}, {Berlin}, {Bernardi}, {Blouin}, {Baudin}, {Bodin}, {Boisnard},
  {Boit}, {Bonneau}, {Borzeix}, {Briet}, {Buey}, {Butler}, {Cailleau},
  {Cautain}, {Chabaud}, {Chaintreuil}, {Chiavassa}, {Costes}, {Cuna Parrho},
  {de Oliveira Fialho}, {Decaudin}, {Defise}, {Djalal}, {Epstein}, {Exil},
  {Faur{\'e}}, {Fenouillet}, {Gaboriaud}, {Gallic}, {Gamet}, {Gavalda},
  {Grolleau}, {Gruneisen}, {Gueguen}, {Guis}, {Guivarc'h}, {Guterman},
  {Hallouard}, {Hasiba}, {Heuripeau}, {Huntzinger}, {Hustaix}, {Imad},
  {Imbert}, {Johlander}, {Jouret}, {Journoud}, {Karioty}, {Kerjean},
  {Lafaille}, {Lafond}, {Lam-Trong}, {Landiech}, {Lapeyrere}, {Larqu{\'e}},
  {Laudet}, {Lautier}, {Lecann}, {Lefevre}, {Leruyet}, {Levacher}, {Magnan},
  {Mazy}, {Mertens}, {Mesnager}, {Meunier}, {Michel}, {Monjoin}, {Naudet},
  {Nguyen-Kim}, {Orcesi}, {Ottacher}, {Perez}, {Peter}, {Plasson}, {Plesseria},
  {Pontet}, {Pradines}, {Quentin}, {Reynaud}, {Rolland}, {Rollenhagen},
  {Romagnan}, {Russ}, {Schmidt}, {Schwartz}, {Sebbag}, {Sedes}, {Smit},
  {Steller}, {Sunter}, {Surace}, {Tello}, {Tiph{\`e}ne}, {Toulouse}, {Ulmer},
  {Vandermarcq}, {Vergnault}, {Vuillemin}, \& {Zanatta}}]{Leger09}
{L{\'e}ger}, A., {Rouan}, D., {Schneider}, J., {et~al.} 2009, \aap, 506, 287

\bibitem[{{Lopez}(2016)}]{Lopez16}
{Lopez}, E.~D. 2016, arXiv:1610.01170, arXiv:1610.01170

\bibitem[{{Lopez} \& {Fortney}(2013)}]{Lopez13}
{Lopez}, E.~D., \& {Fortney}, J.~J. 2013, \apj, 776, 2

\bibitem[{{Lopez} \& {Fortney}(2014)}]{Lopez14}
---. 2014, \apj, 792, 1

\bibitem[{{Lucy} \& {Sweeney}(1971)}]{Lucy71}
{Lucy}, L.~B., \& {Sweeney}, M.~A. 1971, \aj, 76, 544

\bibitem[{{Lundkvist} {et~al.}(2016){Lundkvist}, {Kjeldsen}, {Albrecht},
  {Davies}, {Basu}, {Huber}, {Justesen}, {Karoff}, {Silva Aguirre}, {Van
  Eylen}, {Vang}, {Arentoft}, {Barclay}, {Bedding}, {Campante}, {Chaplin},
  {Christensen-Dalsgaard}, {Elsworth}, {Gilliland}, {Handberg}, {Hekker},
  {Kawaler}, {Lund}, {Metcalfe}, {Miglio}, {Rowe}, {Stello}, {Tingley}, \&
  {White}}]{Lundkvist16}
{Lundkvist}, M.~S., {Kjeldsen}, H., {Albrecht}, S., {et~al.} 2016, Nature
  Communications, 7, 11201

\bibitem[{{Mann} {et~al.}(2017){Mann}, {Gaidos}, {Vanderburg}, {Rizzuto},
  {Ansdell}, {Medina}, {Mace}, {Kraus}, \& {Sokal}}]{Mann17}
{Mann}, A.~W., {Gaidos}, E., {Vanderburg}, A., {et~al.} 2017, \aj, 153, 64

\bibitem[{{Marcus} {et~al.}(2010){Marcus}, {Sasselov}, {Hernquist}, \&
  {Stewart}}]{Marcus10}
{Marcus}, R.~A., {Sasselov}, D., {Hernquist}, L., \& {Stewart}, S.~T. 2010,
  \apjl, 712, L73

\bibitem[{{Marcy} \& {Butler}(1992)}]{Marcy92}
{Marcy}, G.~W., \& {Butler}, R.~P. 1992, \pasp, 104, 270

\bibitem[{{Marcy} {et~al.}(2014){Marcy}, {Isaacson}, {Howard}, {Rowe},
  {Jenkins}, {Bryson}, {Latham}, {Howell}, {Gautier}, {Batalha}, {Rogers},
  {Ciardi}, {Fischer}, {Gilliland}, {Kjeldsen}, {Christensen-Dalsgaard},
  {Huber}, {Chaplin}, {Basu}, {Buchhave}, {Quinn}, {Borucki}, {Koch}, {Hunter},
  {Caldwell}, {Van Cleve}, {Kolbl}, {Weiss}, {Petigura}, {Seager}, {Morton},
  {Johnson}, {Ballard}, {Burke}, {Cochran}, {Endl}, {MacQueen}, {Everett},
  {Lissauer}, {Ford}, {Torres}, {Fressin}, {Brown}, {Steffen}, {Charbonneau},
  {Basri}, {Sasselov}, {Winn}, {Sanchis-Ojeda}, {Christiansen}, {Adams},
  {Henze}, {Dupree}, {Fabrycky}, {Fortney}, {Tarter}, {Holman}, {Tenenbaum},
  {Shporer}, {Lucas}, {Welsh}, {Orosz}, {Bedding}, {Campante}, {Davies},
  {Elsworth}, {Handberg}, {Hekker}, {Karoff}, {Kawaler}, {Lund}, {Lundkvist},
  {Metcalfe}, {Miglio}, {Silva Aguirre}, {Stello}, {White}, {Boss}, {Devore},
  {Gould}, {Prsa}, {Agol}, {Barclay}, {Coughlin}, {Brugamyer}, {Mullally},
  {Quintana}, {Still}, {Thompson}, {Morrison}, {Twicken}, {D{\'e}sert},
  {Carter}, {Crepp}, {H{\'e}brard}, {Santerne}, {Moutou}, {Sobeck}, {Hudgins},
  {Haas}, {Robertson}, {Lillo-Box}, \& {Barrado}}]{Marcy14a}
{Marcy}, G.~W., {Isaacson}, H., {Howard}, A.~W., {et~al.} 2014, \apjs, 210, 20

\bibitem[{{Meunier} {et~al.}(2010){Meunier}, {Desort}, \&
  {Lagrange}}]{Meunier10}
{Meunier}, N., {Desort}, M., \& {Lagrange}, A.-M. 2010, \aap, 512, A39

\bibitem[{{Middelkoop}(1982)}]{Middelkoop82}
{Middelkoop}, F. 1982, \aap, 107, 31

\bibitem[{{Morton}(2015)}]{Morton15a}
{Morton}, T.~D. 2015, {isochrones: Stellar model grid package}, Astrophysics
  Source Code Library, ascl:1503.010

\bibitem[{{Nelson} {et~al.}(2014){Nelson}, {Ford}, {Wright}, {Fischer}, {von
  Braun}, {Howard}, {Payne}, \& {Dindar}}]{Nelson14}
{Nelson}, B.~E., {Ford}, E.~B., {Wright}, J.~T., {et~al.} 2014, \mnras, 441,
  442

\bibitem[{{Noyes} {et~al.}(1984){Noyes}, {Hartmann}, {Baliunas}, {Duncan}, \&
  {Vaughan}}]{Noyes84}
{Noyes}, R.~W., {Hartmann}, L.~W., {Baliunas}, S.~L., {Duncan}, D.~K., \&
  {Vaughan}, A.~H. 1984, \apj, 279, 763

\bibitem[{{O'Toole} {et~al.}(2009){O'Toole}, {Tinney}, {Jones}, {Butler},
  {Marcy}, {Carter}, \& {Bailey}}]{Otoole09}
{O'Toole}, S.~J., {Tinney}, C.~G., {Jones}, H.~R.~A., {et~al.} 2009, \mnras,
  392, 641

\bibitem[{{Owen} \& {Wu}(2013)}]{Owen13}
{Owen}, J.~E., \& {Wu}, Y. 2013, \apj, 775, 105

\bibitem[{{Pepe} {et~al.}(2013){Pepe}, {Cameron}, {Latham}, {Molinari}, {Udry},
  {Bonomo}, {Buchhave}, {Charbonneau}, {Cosentino}, {Dressing}, {Dumusque},
  {Figueira}, {Fiorenzano}, {Gettel}, {Harutyunyan}, {Haywood}, {Horne},
  {Lopez-Morales}, {Lovis}, {Malavolta}, {Mayor}, {Micela}, {Motalebi},
  {Nascimbeni}, {Phillips}, {Piotto}, {Pollacco}, {Queloz}, {Rice}, {Sasselov},
  {S{\'e}gransan}, {Sozzetti}, {Szentgyorgyi}, \& {Watson}}]{Pepe13}
{Pepe}, F., {Cameron}, A.~C., {Latham}, D.~W., {et~al.} 2013, \nat, 503, 377

\bibitem[{{Petigura}(2015)}]{Petigura15b}
{Petigura}, E.~A. 2015, PhD thesis, University of California, Berkeley,
  arXiv:1510.03902

\bibitem[{{Petigura} {et~al.}(2013){Petigura}, {Marcy}, \&
  {Howard}}]{Petigura13b}
{Petigura}, E.~A., {Marcy}, G.~W., \& {Howard}, A.~W. 2013, \apj, 770, 69

\bibitem[{{Petigura} {et~al.}(2017){Petigura}, {Sinukoff}, {Lopez},
  {Crossfield}, {Howard}, {Brewer}, {Fulton}, {Isaacson}, {Ciardi}, {Howell},
  {Everett}, {Horch}, {Hirsch}, {Weiss}, \& {Schlieder}}]{Petigura17}
{Petigura}, E.~A., {Sinukoff}, E., {Lopez}, E., {et~al.} 2017, ArXiv e-prints,
  arXiv:1702.00013

\bibitem[{{Pope} {et~al.}(2016){Pope}, {Parviainen}, \& {Aigrain}}]{Pope16}
{Pope}, B.~J.~S., {Parviainen}, H., \& {Aigrain}, S. 2016, \mnras, 461, 3399

\bibitem[{Powell(1964)}]{Powell64}
Powell, M. J.~D. 1964, The Computer Journal, 7, 155

\bibitem[{{Rogers}(2015)}]{Rogers15}
{Rogers}, L.~A. 2015, \apj, 801, 41

\bibitem[{{Rogers} \& {Seager}(2010)}]{Rogers10}
{Rogers}, L.~A., \& {Seager}, S. 2010, \apj, 712, 974

\bibitem[{{Sanchis-Ojeda} {et~al.}(2014){Sanchis-Ojeda}, {Rappaport}, {Winn},
  {Kotson}, {Levine}, \& {El Mellah}}]{Ojeda14}
{Sanchis-Ojeda}, R., {Rappaport}, S., {Winn}, J.~N., {et~al.} 2014, \apj, 787,
  47

\bibitem[{{Sanchis-Ojeda} {et~al.}(2013){Sanchis-Ojeda}, {Rappaport}, {Winn},
  {Levine}, {Kotson}, {Latham}, \& {Buchhave}}]{Ojeda13}
---. 2013, \apj, 774, 54

\bibitem[{{Sinukoff} {et~al.}(2016){Sinukoff}, {Howard}, {Petigura},
  {Schlieder}, {Crossfield}, {Ciardi}, {Fulton}, {Isaacson}, {Aller},
  {Baranec}, {Beichman}, {Hansen}, {Knutson}, {Law}, {Liu}, {Riddle}, \&
  {Dressing}}]{Sinukoff16}
{Sinukoff}, E., {Howard}, A.~W., {Petigura}, E.~A., {et~al.} 2016, \apj, 827,
  78

\bibitem[{{Sinukoff} {et~al.}(2017){Sinukoff}, {Howard}, {Petigura}, {Fulton},
  {Isaacson}, {Weiss}, {Brewer}, {Hansen}, {Hirsch}, {Christiansen}, {Crepp},
  {Crossfield}, {Schlieder}, {Ciardi}, {Beichman}, {Knutson}, {Benneke},
  {Dressing}, {Livingston}, {Deck}, {L{\'e}pine}, \& {Rogers}}]{Sinukoff17a}
---. 2017, \aj, 153, 70

\bibitem[{{Valencia} {et~al.}(2013){Valencia}, {Guillot}, {Parmentier}, \&
  {Freedman}}]{Valencia13}
{Valencia}, D., {Guillot}, T., {Parmentier}, V., \& {Freedman}, R.~S. 2013,
  \apj, 775, 10

\bibitem[{{Valenti} {et~al.}(1995){Valenti}, {Butler}, \& {Marcy}}]{Valenti95}
{Valenti}, J.~A., {Butler}, R.~P., \& {Marcy}, G.~W. 1995, \pasp, 107, 966

\bibitem[{{Valenti} \& {Fischer}(2005)}]{Valenti05}
{Valenti}, J.~A., \& {Fischer}, D.~A. 2005, \apjs, 159, 141

\bibitem[{{Valsecchi} {et~al.}(2015){Valsecchi}, {Rappaport}, {Rasio},
  {Marchant}, \& {Rogers}}]{Valsecchi15}
{Valsecchi}, F., {Rappaport}, S., {Rasio}, F.~A., {Marchant}, P., \& {Rogers},
  L.~A. 2015, \apj, 813, 101

\bibitem[{{Valsecchi} {et~al.}(2014){Valsecchi}, {Rasio}, \&
  {Steffen}}]{Valsecchi14}
{Valsecchi}, F., {Rasio}, F.~A., \& {Steffen}, J.~H. 2014, \apjl, 793, L3

\bibitem[{{Vanderburg} {et~al.}(2015){Vanderburg}, {Montet}, {Johnson},
  {Buchhave}, {Zeng}, {Pepe}, {Collier Cameron}, {Latham}, {Molinari}, {Udry},
  {Lovis}, {Matthews}, {Cameron}, {Law}, {Bowler}, {Angus}, {Baranec},
  {Bieryla}, {Boschin}, {Charbonneau}, {Cosentino}, {Dumusque}, {Figueira},
  {Guenther}, {Harutyunyan}, {Hellier}, {Kuschnig}, {Lopez-Morales}, {Mayor},
  {Micela}, {Moffat}, {Pedani}, {Phillips}, {Piotto}, {Pollacco}, {Queloz},
  {Rice}, {Riddle}, {Rowe}, {Rucinski}, {Sasselov}, {S{\'e}gransan},
  {Sozzetti}, {Szentgyorgyi}, {Watson}, \& {Weiss}}]{Vanderburg15}
{Vanderburg}, A., {Montet}, B.~T., {Johnson}, J.~A., {et~al.} 2015, \apj, 800,
  59

\bibitem[{{Vanderburg} {et~al.}(2016){Vanderburg}, {Bieryla}, {Duev},
  {Jensen-Clem}, {Latham}, {Mayo}, {Baranec}, {Berlind}, {Kulkarni}, {Law},
  {Nieberding}, {Riddle}, \& {Salama}}]{Vanderburg16}
{Vanderburg}, A., {Bieryla}, A., {Duev}, D.~A., {et~al.} 2016, \apjl, 829, L9

\bibitem[{{Vogt} {et~al.}(1994){Vogt}, {Allen}, {Bigelow}, {Bresee}, {Brown},
  {Cantrall}, {Conrad}, {Couture}, {Delaney}, {Epps}, {Hilyard}, {Hilyard},
  {Horn}, {Jern}, {Kanto}, {Keane}, {Kibrick}, {Lewis}, {Osborne},
  {Pardeilhan}, {Pfister}, {Ricketts}, {Robinson}, {Stover}, {Tucker}, {Ward},
  \& {Wei}}]{Vogt94}
{Vogt}, S.~S., {Allen}, S.~L., {Bigelow}, B.~C., {et~al.} 1994, in Society of
  Photo-Optical Instrumentation Engineers (SPIE) Conference Series, Vol. 2198,
  Instrumentation in Astronomy VIII, ed. D.~L. {Crawford} \& E.~R. {Craine},
  362

\bibitem[{{von Braun} {et~al.}(2011){von Braun}, {Boyajian}, {ten Brummelaar},
  {Kane}, {van Belle}, {Ciardi}, {Raymond}, {L{\'o}pez-Morales}, {McAlister},
  {Schaefer}, {Ridgway}, {Sturmann}, {Sturmann}, {White}, {Turner},
  {Farrington}, \& {Goldfinger}}]{VonBraun11}
{von Braun}, K., {Boyajian}, T.~S., {ten Brummelaar}, T.~A., {et~al.} 2011,
  \apj, 740, 49

\bibitem[{{Weiss} \& {Marcy}(2014)}]{Weiss14}
{Weiss}, L.~M., \& {Marcy}, G.~W. 2014, \apjl, 783, L6

\bibitem[{{Winn} {et~al.}(2017){Winn}, {Sanchis-Ojeda}, {Rogers}, {Petigura},
  {Howard}, {Isaacson}, {Marcy}, \& {Schlaufman}}]{Winn17}
{Winn}, J.~N., {Sanchis-Ojeda}, R., {Rogers}, L., {et~al.} 2017,
  arXiv:1704.00203

\bibitem[{{Winn} {et~al.}(2010){Winn}, {Johnson}, {Howard}, {Marcy}, {Bakos},
  {Hartman}, {Torres}, {Albrecht}, \& {Narita}}]{Winn10}
{Winn}, J.~N., {Johnson}, J.~A., {Howard}, A.~W., {et~al.} 2010, \apj, 718, 575

\end{thebibliography}
\end{document}